\documentclass[reqno,draft]{amsart}
\usepackage{amssymb}
\usepackage{upref}

\newcommand{\bbN}{{\mathbb{N}}}
\newcommand{\bbR}{{\mathbb{R}}}

\newcommand{\bbZ}{{\mathbb{Z}}}
\newcommand{\bbC}{{\mathbb{C}}}
\newcommand{\bbT}{{\mathbb{T}}}

\newcommand{\calD}{{\mathcal D}}

\newcommand{\calF}{{\mathcal F}}

\newcommand{\calK}{{\mathcal K}}

\newcommand{\Green}{{\mathcal G}}

\newcommand{\dott}{\,\cdot\,}

\newcommand{\no}{\nonumber}
\newcommand{\lb}{\label}
\newcommand{\f}{\frac}
\newcommand{\ul}{\underline}
\newcommand{\ol}{\overline}
\newcommand{\ti}{\tilde}
\newcommand{\wti}{\widetilde}

\newcommand{\Oh}{O}
\newcommand{\oh}{o}

\newcommand{\loc}{\text{\rm{loc}}}

\newcommand{\bi}{\bibitem}

\newcommand{\Pinfp}{{P_{\infty_+}}}
\newcommand{\Pinfm}{{P_{\infty_-}}}

\newcommand{\Pinfpm}{{P_{\infty_\pm}}}

\renewcommand{\Re}{\text{\rm Re}}
\renewcommand{\Im}{\text{\rm Im}}

\DeclareMathOperator{\CH}{CH}
\DeclareMathOperator{\sCH}{s-CH}

\allowdisplaybreaks
\numberwithin{equation}{section}

\newtheorem{theorem}{Theorem}[section]
\newtheorem{lemma}[theorem]{Lemma}
\newtheorem{corollary}[theorem]{Corollary}
\newtheorem{hypothesis}[theorem]{Hypothesis}
\newtheorem{remark}[theorem]{Remark}
\theoremstyle{definition}
\newtheorem{definition}[theorem]{Definition}

\newcommand{\abs}[1]{\lvert#1\rvert}

\begin{document}
\title[Real algebro-geometric solutions of the CH
hierarchy]{Real-Valued Algebro-Geometric Solutions \\ of the
Camassa--Holm hierarchy}
\author[F.\ Gesztesy]{Fritz Gesztesy}
\address{Department of Mathematics,
University of Missouri,
Columbia, MO 65211, USA}
\email{fritz@math.missouri.edu}
\urladdr{http://www.math.missouri.edu/people/fgesztesy.html}
\author[H.\ Holden]{Helge Holden}
\address{Department of Mathematical Sciences,
Norwegian University of
Science and Technology, NO--7491 Trondheim, Norway}
\email{holden@math.ntnu.no}
\urladdr{http://www.math.ntnu.no/\~{}holden/}
\thanks{Research supported in part by the Research Council of Norway.}
\date{August 14, 2002}

\begin{abstract}
We provide a treatment of real-valued, smooth, and bounded
algebro-geometric solutions of the Camassa--Holm (CH) hierarchy and
describe the associated isospectral torus. We also discuss
real-valued algebro-geometric solutions with a cusp behavior.
\end{abstract}

\maketitle

\section{Introduction}\lb{s1}

Recently, we provided a detailed treatment of the Camassa--Holm (CH) 
hierarchy with special emphasis on its algebro-geometric solutions. The
first nonlinear partial differential equation of this hierarchy, the
Camassa--Holm equation, also known as the dispersive shallow water
equation 
\cite{CamassaHolm:1993} is given by 
\begin{equation}
4u_t-u_{xxt}-2u u_{xxx}-4u_x u_{xx}+24u u_x=0, \quad (x,t)\in\bbR^2
\lb{ch1.1}
\end{equation}
(choosing a convenient scaling of $x,t$). For various aspects of local
and global existence,  and uniqueness of  solutions of
\eqref{ch1.1}, wave breaking phenomena, soliton-type solutions 
(``peakons''), complete  integrability aspects such as infinitely many
conservation laws, (bi-)Hamiltonian formalism, B\"acklund
transformations, infinite dimensional symmetry groups, etc., we refer to
the literature provided in \cite{GH02} (see also 
\cite[Ch.\ 5]{GesztesyHolden:2000}). The case of spatially periodic
solutions, the corresponding inverse spectral problem, isospectral
classes of solutions, and quasi-periodicity of solutions with respect to
time are discussed in \cite{Co97}--\cite{ConstantinMcKean:1999}. Moreover,
algebro-geometric solutions of \eqref{ch1.1} and their properties are
studied in 
\cite{AlberCamassaFedorovHolmMarsden:2001}--\cite{AlberFedorov:2001}, and 
\cite{GH02} (see also \cite[Ch.\ 5]{GesztesyHolden:2000}).

In Section \ref{s2} we recall the basic polynomial recursion formalism
that  defines the CH hierarchy using a zero-curvature approach. Section 
\ref{s3} recalls the stationary CH hierarchy and the associated 
algebro-geometric formalism. Section \ref{s4} provides a brief summary of
self-adjoint canonical systems as needed in this paper, and Section
\ref{s5} finally discusses the principal result of this paper, the class
of real-valued, smooth, and bounded  algebro-geometric solutions of the CH
hierarchy and the associated isospectral torus. 

This paper should be viewed as a companion to our paper \cite{GH02} (see
also \cite[Ch.\ 5]{GesztesyHolden:2000}) on the CH hierarchy and we refer
to it for background material and pertinent references on the subject.

\section[Fundamentals of the CH hierarchy]{The CH hierarchy,
recursion relations, and hyperelliptic curves}  \label{s2}

In this section we review the basic construction of the Camassa--Holm
hierarchy using  a zero-curvature approach following \cite{GH02} (see also 
\cite[Ch.\ 5]{GesztesyHolden:2000}). 

Throughout this section we will suppose the following hypothesis
($\bbN_0=\bbN\cup\{0\}$).
\begin{hypothesis}\lb{hypo-ch1.1}
In the stationary case we assume that
\begin{equation}
u\in C^\infty(\bbR), \; \f{d^m u}{dx^m}\in L^\infty(\bbR), \; 
m\in\bbN_{0}. \lb{ch2.1}
\end{equation}
In the time-dependent case $($cf.\ \eqref{ch2.40}--\eqref{ch2.47}$)$ we
suppose 
\begin{align}
&u(\dott,t)\in C^\infty(\bbR), \; \f{\partial^m u}{\partial
x^m}(\dott,t)\in L^\infty(\bbR), \;  m\in\bbN_{0}, \, t\in\bbR, \no \\
&u(x,\dott), u_{xx}(x,\dott)\in C^1(\bbR), \; x\in\bbR. \lb{ch2.2}
\end{align}
\end{hypothesis}

We start by formulating the basic polynomial setup. One defines
$\{f_\ell\}_{\ell\in\bbN_0}$ recursively by
\begin{equation}
f_{0}=1, \quad f_{\ell,x}=-2\Green\big(2(4u-u_{xx})f_{\ell -1,x}
+(4u_{x}-u_{xxx})f_{\ell -1}  \big), \quad \ell\in\bbN, \lb{ch2.3} 
\end{equation}
where $\Green$ is given by
\begin{equation}
\Green\colon L^\infty(\bbR)\to L^\infty(\bbR), \quad
(\Green v)(x)=\frac14 \int_{\bbR} dy\, e^{-2\abs{x-y}} v(y),
\quad x\in\bbR, \; v\in L^\infty(\bbR). \lb{ch2.4} 
\end{equation}
One observes that $\Green$ is the 
resolvent of minus the one-dimensional Laplacian at energy parameter
equal to $-4$, that is,
\begin{equation}
    \Green=\bigg(-\frac{d^2}{dx^2}+4\bigg)^{-1}. \lb{ch2.5} 
\end{equation}
The first coefficient reads
\begin{equation}
f_1=-2u+c_1,\lb{ch2.6} 
\end{equation}
where $c_1$ is an integration constant.  Subsequent coefficients
are non-local with respect to $u$. At each level a new integration
constant,  denoted by $c_{\ell}$, is introduced.
Moreover, we introduce coefficients  
$\{g_\ell\}_{\ell\in\bbN_0}$ and $\{h_\ell\}_{\ell\in\bbN_0}$ by
\begin{equation}
g_\ell=f_{\ell}+ \f12 f_{\ell,x}, \quad 
h_{\ell} = (4u-u_{xx})f_{\ell} - g_{\ell+1,x}, 
\quad \ell\in\bbN_{0}. \lb{ch2.8} 
\end{equation}
Explicitly, one computes
\begin{align}
f_{0}&=1, \quad 
f_{1}=-2u+c_{1}, \quad
f_{2}=2u^2+2\Green\big(u_{x}^2+8u^2\big)+c_1(-2 u)+c_2, 
\no \\
g_{0}&=1,\quad g_{1}=-2u-u_{x}+c_{1}, \lb{ch2.9} \\
g_{2}&=2u^2+2uu_x+2\Green\big(u_x^2
+ u_x u_{xx}+8u u_x+8u^2\big) +c_1(-2u-u_x)+c_2, \no \\ 
h_{0}&=4u+2u_{x}, \no\\
h_{1}&=-2u_x^2-4u u_{x}-8u^2 \no \\
&\quad-2\Green\big(u_x u_{xxx}+u_{xx}^2+2u_x u_{xx}+8uu_{xx}
+8u_x^2+16uu_x \big) \no \\
& \quad +c_1(4u+2u_{x}), \, \text{  etc.}\no
\end{align}

Given Hypothesis \ref{hypo-ch1.1}, one introduces the $2\times 2$ 
matrix $U$ by
\begin{align}
U(z,x)&=\begin{pmatrix}-1 &1\\
z^{-1}(4u(x)-u_{xx}(x)) &1 \end{pmatrix}, \quad 
x\in\bbR, \lb{ch2.14} 
\end{align}
and for each $n\in\bbN_{0}$ the following $2\times 2$  matrix $V_n$ by
\begin{equation}
V_{n}(z,x)=
\begin{pmatrix}-G_{n}(z,x)& F_{n}(z,x)\\
    z^{-1} H_{n}(z,x) &G_{n}(z,x)
\end{pmatrix}, \quad n\in\bbN_0, \, z\in\bbC\backslash\{0\}, 
\, x\in\bbR, \lb{ch2.15} 
\end{equation} 
assuming $F_n$, $G_n$, and $H_n$ to be polynomials of degree $n$ with 
respect to $z$ and $C^\infty$ in $x$. Postulating the zero-curvature 
condition
\begin{equation}
-V_{n,x}(z,x)+[U(z,x),V_n(z,x)]=0, \lb{ch2.16}
\end{equation}
one finds
\begin{align}
F_{n,x}(z,x)&=2 G_{n}(z,x)-2F_n(z,x), \lb{ch2.17} \\
 zG_{n,x}(z,x)&=(4u(x)-u_{xx}(x))F_{n}(z,x)-H_n(z,x), \lb{ch2.18}
\\ H_{n,x}(z,x)&=2H_{n}(z,x)-2(4u(x)-u_{xx}(x)) G_n(z,x).
\lb{ch2.19}
\end{align}
{}From \eqref{ch2.17}--\eqref{ch2.19} one infers that
\begin{equation}
\f{d}{dx}\det(V_n(z,x))=-\f{1}{z}\f{d}{dx}\bigg(zG_{n}(z,x)^2+F_n(z,x)
H_n(z,x)\bigg)=0, \lb{ch2.20}
\end{equation}
and hence
\begin{equation}
z^2G_{n}(z,x)^2+zF_n(z,x) H_n(z,x)=R_{2n+2}(z), \lb{ch2.23}
\end{equation}
where the polynomial $R_{2n+2}$ of degree $2n+2$ is $x$-independent,
\begin{equation} 
R_{2n+2}(z)=\prod_{m=0}^{2n+1}(z-E_m), \quad E_0,  
E_1,\dots,E_{2n}\in\bbC, \; E_{2n+1}=0. \lb{ch2.22}
\end{equation}

Next one makes the ansatz that $F_{n}$, $H_{n}$, and $G_{n}$ are
polynomials of degree $n$, related to the coefficients 
$f_\ell$, $h_\ell$, and $g_\ell$ by
\begin{equation}
F_{n}(z,x)=\sum_{\ell=0}^n f_{n-\ell}(x)z^\ell, \quad
G_{n}(z,x)=\sum_{\ell=0}^{n}g_{n-\ell}(x)z^\ell, \quad 
H_{n}(z,x)=\sum_{\ell=0}^n h_{n-\ell}(x)z^\ell. \lb{ch2.26}
\end{equation}
Insertion of \eqref{ch2.26} into \eqref{ch2.17}--\eqref{ch2.19} then
yields the recursion relations \eqref{ch2.3}, \eqref{ch2.4} and
\eqref{ch2.8} for $f_{\ell}$ and $g_{\ell}$ for $\ell=0,\dots,n$. For fixed
$n\in\bbN$ we  obtain  the recursion \eqref{ch2.8} for $h_{\ell}$ for
$\ell=0,\dots,n-1$  and 
\begin{equation}
    h_{n}=(4u-u_{xx})f_{n}. \lb{ch2.27}
\end{equation}
(When $n=0$ one directly gets $h_{0}=(4u-u_{xx})$.) Moreover,
taking $z=0$ in \eqref{ch2.23} yields
\begin{equation}
f_n(x)h_n(x)=-\prod_{m=0}^{2n} E_m. \lb{ch2.27a} 
\end{equation}
In addition, one finds
\begin{equation}
    h_{n,x}(x)-2h_{n}(x)+2(4u(x)-u_{xx}(x))g_{n}(x)=0, 
    \quad n\in\bbN_{0}. \lb{ch2.28}
\end{equation}
Using the relations \eqref{ch2.27} and \eqref{ch2.8} permits one to write
\eqref{ch2.28} as
\begin{align}
\sCH_{n}(u)=(u_{xxx}-4u_{x})f_{n}-2(4u-u_{xx})f_{n,x}=0, 
    \quad n\in\bbN_{0}. \lb{ch2.29}
\end{align}
Varying $n\in\bbN_0$ in \eqref{ch2.29} then defines the stationary CH hierarchy. 
We record the first few equations explicitly,
\begin{align}
\sCH_0(u)&=u_{xxx}-4u_{x}=0, \no \\
\sCH_1(u)&=-2u u_{xxx}-4u_{x}u_{xx}+24u u_{x}+ c_{1}(u_{xxx}-4u_{x})=0,
\lb{ch2.30} \\
\sCH_2(u)&=2u^2 u_{xxx}-8u u_{x} u_{xx}-40u^2 u_{x} 
+2(u_{xxx}-4u_x)\Green\big(u_{x}^2+8u^2\big) \no \\ 
&\quad
-8(4u-u_{xx})\Green\big(u_{x} u_{xx}+8u u_{x}\big)
\no \\ &\quad +c_1(-2u u_{xxx}-4u_{x}u_{xx}+24u u_{x})
+c_{2}(u_{xxx}-4u_{x})=0,  
\, \text{  etc.} \no
\end{align}

By definition, the set of solutions of \eqref{ch2.29}, with $n$ ranging in
$\bbN_0$, represents the class of algebro-geometric CH
solutions. If $u$ satisfies one of the stationary CH equations in
\eqref{ch2.29} for a particular value of $n$, then it satisfies 
infinitely many
such equations of order higher than $n$ for certain choices of integration
constants $c_\ell$. At times it will be convenient to abbreviate
(algebro-geometric) stationary CH solutions $u$ simply as CH {\it
potentials}.

Using equations \eqref{ch2.17}--\eqref{ch2.19} one can also derive
individual differential equations for $F_n$ and $H_n$. Focusing on
$F_n$ only, one obtains
\begin{align}
&F_{n,xxx}(z,x)-4\big(z^{-1}(4u(x)-u_{xx}(x))+1\big)F_{n,x}(z,x)
\no \\
& \quad -2z^{-1}(4u_{x}(x)-u_{xxx}(x))F_{n}(z,x)=0 \lb{ch2.37}    
\end{align}
and
\begin{align}
&-(z^2/2)F_{n,xx}(z,x)F_n(z,x)+(z^2/4)F_{n,x}(z,x)^2+z^2 F_n(z,x)^2 
\no \\
& \quad +z(4u(x)-u_{xx}(x))F_n(z,x)^2=R_{2n+2}(z). \lb{ch2.39a}
\end{align}
Equation \eqref{ch2.39a} leads to an explicit determination
of the integration constants $c_1,\dots,c_{n}$ in the stationary CH
equations \eqref{ch2.29} in terms of the zeros $E_0=0$,
$E_1,\dots,E_{2n+1}$ of the associated polynomial $R_{2n+2}$ in
\eqref{ch2.22}. In fact, one can prove 
\begin{equation}
c_\ell=c_\ell(\ul E), \quad \ell=0,\dots,n, \lb{ch2.39C}
\end{equation}
where 
\begin{align}
c_0(\ul E)&=1,\no \\
c_k(\ul E)&=-\!\!\!\!\!\sum_{\substack{j_1,\dots,j_{2n+1}=0\\
 j_1+\cdots+j_{2n+1}=k}}^{k}\!\!
\f{(2j_1)!\cdots(2j_{2n+1})!}
{2^{2k} (j_1!)^2\cdots (j_{2n+1}!)^2 (2j_1-1)\cdots(2j_{2n+1}-1)} \no \\
& \hspace*{2.4cm} \times E_1^{j_1}\cdots E_{2n+1}^{j_{2n+1}}, \quad 
k\in\bbN. \label{ch2.39D}
\end{align}

Next, we turn to the time-dependent CH hierarchy. Introducing a    
deformation parameter $t_n\in\bbR$ into $u$ (i.e., replacing 
$u(x)$ by $u(x,t_n)$), the definitions \eqref{ch2.14}, 
\eqref{ch2.15}, and \eqref{ch2.26} of $U$, $V_n$, and 
$F_n$, $G_n$,
and $H_n$, respectively, still apply. The corresponding zero-curvature 
relation reads
\begin{equation}
U_{t_n}(z,x,t_n)-V_{n,x}(z,x,t_n)+[U(z,x,t_n),V_n(z,x,t_n)]=0, 
\quad n\in\bbN_0,\lb{ch2.40}
\end{equation}
which results in the following set of time-dependent equations
\begin{align}
&4u_{t_n}(x,t_n)-u_{xxt_{n}}(x,t_n)-
H_{n,x}(z,x,t_{n})+2H_{n}(z,x,t_{n}) \no \\
&\quad -2(4u(x,t_n)-u_{xx}(x,t_n))G_{n}(z,x,t_{n})=0, \lb{ch2.41} \\
&F_{n,x}(z,x,t_n)=2G_{n}(z,x,t_n)-2F_n(z,x,t_n), \lb{ch2.42} \\
&zG_{n,x}(z,x,t_n)=(4u(x,t_n)-u_{xx}(x,t_n))F_n(z,x,t_n)
-H_{n}(z,x,t_n).  \lb{ch2.43}
\end{align}
Inserting the polynomial expressions for $F_n$, $H_n$,
and $G_{n}$ into \eqref{ch2.42} and  \eqref{ch2.43}, respectively, first
yields  recursion  relations \eqref{ch2.3} and \eqref{ch2.8} for
$f_{\ell}$ and $g_{\ell}$ for $\ell=0,\dots,n$. For fixed $n\in\bbN$ we 
obtain {}from \eqref{ch2.41} the recursion for $h_{\ell}$ for
$\ell=0,\dots,n-1$ and 
\begin{equation}
    h_{n}=(4u-u_{xx})f_{n}. \lb{ch2.44}
\end{equation}
(When $n=0$ one directly gets $h_{0}=(4u-u_{xx})$.)
In addition, one finds
\begin{align}
&4u_{t_n}(x,t_n)-u_{xxt_n}(x,t_n)-h_{n,x}(x,t_n)+2h_{n}(x,t_n)\no \\
&\quad -2(4u(x,t_n)-u_{xx}(x,t_n))g_{n}(x,t_n)=0, 
    \quad n\in\bbN_{0}. \lb{ch2.45}
\end{align}
Using relations \eqref{ch2.27} and \eqref{ch2.44} permits one to write 
\eqref{ch2.45} as
\begin{align}
\CH_{n}(u)=4u_{t_n}-u_{xxt_n}+(u_{xxx}-4u_{x})f_{n}-2(4u-u_{xx})f_{n,x}=0,
\quad n\in\bbN_{0}. \lb{ch2.46}
\end{align}
Varying $n\in\bbN_0$ in \eqref{ch2.46} then defines the time-dependent 
CH hierarchy. We record the first few equations explicitly,
\begin{align}
\CH_0(u)&=4u_{t_{0}}-u_{xxt_{0}}+u_{xxx}-4u_{x}=0, \no \\
\CH_1(u)&=4u_{t_{1}}-u_{xxt_{1}} -2u u_{xxx}-4u_{x}u_{xx}
+24u u_{x}+c_{1}(u_{xxx}-4u_{x})=0, \no \\
\CH_2(u)&=4u_{t_{2}}-u_{xxt_{2}}+2u^2 u_{xxx}-8u u_{x} u_{xx}
-40u^2 u_{x} \lb{ch2.47} \\
& \quad +2(u_{xxx}-4u_x)\Green\big(u_{x}^2+8u^2\big) 
-8(4u-u_{xx})\Green\big(u_{x} u_{xx}+8u u_{x} \big) \no \\ 
&\quad +c_1(-2u u_{xxx}-4u_{x}u_{xx}+24u u_{x})
+c_{2}(u_{xxx}-4u_{x})=0, \, \text{  etc.} \no
\end{align}
Up to an inessential scaling of the $(x,t_1)$ variables,
${\CH}_1(u)=0$ with $c_1=0$ represents {\it the Camassa--Holm} equation as
discussed in \cite{CamassaHolm:1993}.

\section{The algebro-geometric CH formalism} \lb{s3}

This section is devoted to a quick review of the stationary CH
hierarchies and the corresponding algebro-geometric formalism as derived
in \cite{GH02} (cf.\ also \cite[Ch.\ 5]{GesztesyHolden:2000}). 

We start with the stationary hierarchy and suppose 
\begin{equation}
u\in C^\infty(\bbR), \; \f{d^m u}{dx^m}\in L^\infty(\bbR), \; 
m\in\bbN_{0}, \label{ch3.0}
\end{equation}
and assume  \eqref{ch2.3}, \eqref{ch2.8}, \eqref{ch2.14}--\eqref{ch2.16},  
\eqref{ch2.23}--\eqref{ch2.29}, keeping $n\in\bbN_0$ fixed.

Recalling \eqref{ch2.22},  
\begin{equation}
R_{2n+2}(z)=\prod_{m=0}^{2n+1}(z-E_m), \quad E_0,E_1,\dots,
E_{2n}\in\bbC, \; E_{2n+1}=0, \lb{ch3.1}
\end{equation}
we introduce the (possibly singular)
hyperelliptic  curve $\calK_n$ of arithmetic genus $n$ defined by
\begin{equation}
\calK_n \colon \calF_n(z,y)=y^2-R_{2n+2}(z)=0. \lb{ch3.3}
\end{equation}
In the following we will occasionally impose further constraints on
the zeros $E_m$ of $R_{2n+2}$ introduced in \eqref{ch3.1} and assume that 
\begin{equation}
E_1,\dots,E_{2n}\in\bbC\backslash\{0\}, \; E_{2n+1}=0. \lb{ch3.3a}
\end{equation}
We compactify $\calK_n$ by adding two points at infinity,  $\Pinfp$, 
$\Pinfm$, with  $\Pinfp\neq \Pinfm$, still denoting
its projective closure by  $\calK_n$.  Hence $\calK_n$
becomes a two-sheeted Riemann surface
of arithmetic genus $n$.  Points $P$ on $\calK_n\backslash\{\Pinfpm\}$ are
denoted by $P=(z,y)$, where $y(\dott)$ denotes the meromorphic
function on $\calK_n$ satisfying $\calF_n(z,y)=0$. 

For notational simplicity we will usually tacitly assume that $n\in\bbN$
(the case $n=0$ being trivial).

In the following the roots of the polynomials $F_n$ and $H_n$ will play a
special role and hence we introduce on $\bbC\times\bbR$
\begin{equation}
F_n(z,x)=\prod_{j=1}^n(z-\mu_j(x)), \quad
H_n(z,x)=h_{0}(x)\prod_{j=1}^n(z-\nu_j(x)), \lb{ch3.5}
\end{equation}
temporarily assuming
\begin{equation}
h_0(x)\neq 0, \quad x\in\bbR. \lb{ch3.5a}
\end{equation}
Moreover, we introduce
\begin{align}
\hat\mu_j(x)&=(\mu_j(x),-\mu_j(x)G_{n}(\mu_j(x),x))\in\calK_n,
\quad j=1,\dots,n, \; x\in\bbR, \lb{ch3.6}\\
\hat\nu_j(x)&=(\nu_j(x),\nu_j(x)G_{n}(\nu_j(x),x))\in\calK_n,
\quad j=1,\dots,n, \; x\in\bbR, \lb{ch3.7}
\end{align}
and
\begin{equation}
P_0 = (0,0). \lb{ch3.8}
\end{equation}
The branch of $y(\dott)$ near $\Pinfpm$ is fixed according to
\begin{equation}
\lim_{\substack{\abs{z(P)}\to\infty\\P\to\Pinfpm}}\f{y(P)}{z(P)
G_n(z(P),x)}=\mp 1. \lb{ch3.8a}
\end{equation}
Due to assumption \eqref{ch3.0}, $u$ is smooth and bounded, and hence  
$F_n(z,\dott)$ and $H_{n}(z,\dott)$ share the same property. 
Thus, one concludes 
\begin{equation}
\mu_j, \nu_k \in C(\bbR), \; j,k=1,\dots,n,  
\lb{ch3.9}
\end{equation}
taking multiplicities (and appropriate reordering) of the zeros of $F_n$ 
and $H_n$ into account. 

Next, we introduce the fundamental meromorphic function $\phi(\dott,x)$ on
$\calK_n$ by
\begin{equation}
\phi(P,x)=\f{y-zG_{n}(z,x)}{F_n(z,x)} 
=\f{zH_{n}(z,x)}{y+z G_{n}(z,x)}, \quad
 P=(z,y)\in\calK_n, \, x\in\bbR. \lb{ch3.11}
\end{equation}
Assuming \eqref{ch3.3a} and \eqref{ch3.5a}, the divisor $(\phi(\dott,x))$
of $\phi(\dott,x)$ is given by
\begin{equation}
(\phi(\dott,x))=\calD_{P_0\hat{\ul\nu}(x)}
-\calD_{\Pinfp\hat{\ul\mu}(x)}, \lb{ch3.13}
\end{equation}
taking into account \eqref{ch3.8a}. 
Here we abbreviated
\begin{equation}
\hat{\ul\mu}=\{\hat\mu_1,\dots,\hat\mu_n\}, \, 
\hat{\ul\nu}=\{\hat\nu_1,\dots,\hat\nu_n\} \in \sigma^n\calK_n, \lb{ch3.14}
\end{equation}
where $\sigma^m\calK_n$, $m\in\bbN$, denotes the $m$th symmetric product of
$\calK_n$.  If $h_0$ is permitted to vanish at a point $x_1\in\bbR$, then
for $x=x_1$, the polynomial $H_n(\dott,x_1)$ is at most of degree $n-1$ 
(cf.\ \eqref{ch2.26}) and \eqref{ch3.13} is altered to
\begin{equation}
(\phi(\dott,x_1))=\calD_{P_0\Pinfm\hat\nu_1 (x_1),\dots,
\hat\nu_{n-1}(x_1)}-\calD_{\Pinfp\hat{\ul\mu}(x)}, \lb{ch3.13a}
\end{equation}
that is, one of the $\hat \nu_j (x)$ tends to $\Pinfm$ as $x\to x_1$
(cf.\ also \eqref{ch3.38a}). Analogously one can discuss the case of
several $\hat \nu_j$ approaching $\Pinfm$. Since this can be viewed as
a limiting case of \eqref{ch3.13}, we will henceforth not particularly
distinguish the case $h_0\neq 0$ from the more general situation where
$h_0$ is permitted to vanish.

Given $\phi(\dott,x)$, one defines the associated Baker-Akhiezer vector 
$\Psi(\dott,x,x_0)$ on $\calK_n\backslash\{\Pinfp,\Pinfm,P_0\}$ by
\begin{equation}
\Psi(P,x,x_0)=\begin{pmatrix} \psi_1(P,x,x_0) \\ \psi_2(P,x,x_0)
\end{pmatrix}, \quad
P\in\calK_n\backslash\{\Pinfp,\Pinfm,P_0\}, \; (x,x_0)\in\bbR^2,
\lb{ch3.15}
\end{equation}
where
\begin{align}
\psi_1(P,x,x_0)&=\exp\left(-(1/z) \int_{x_0}^x dx'\,
\phi(P,x')-(x-x_0)\right), \lb{ch3.16}\\
\psi_2(P,x,x_0)&=-\psi_1(P,x,x_0) \phi(P,x)/z. \lb{ch3.17}
\end{align}
Although $\Psi$ is formally the analog of the Baker--Akhiezer
vector of the stationary CH hierarchy when compared to analogous
definitions in the context of the KdV or AKNS hierarchies, its actual
properties in a neighborhood of its essential singularity feature
characteristic differences to standard Baker--Akhiezer vectors as
discussed in \cite[Ch.\ 5]{GesztesyHolden:2000} and \cite{GH02}. 

The basic properties of $\phi$ and $\Psi$ then read as follows.

\begin{lemma} \lb{lemma-ch3.1}
Suppose \eqref{ch3.0}, assume the $n$th stationary CH
equation \eqref{ch2.29} holds, and let $P=(z,y)\in\calK_n\backslash
\{\Pinfp,\Pinfm,P_0\}$, $(x,x_0)\in\bbR^2$. Then
$\phi$ satisfies the Riccati-type equation
\begin{equation}
\phi_x(P,x)-z^{-1}\phi(P,x)^2-2\phi(P,x)+4u(x)-u_{xx}(x)=0, 
\lb{ch3.18}
\end{equation}
as well as
\begin{align}
\phi(P,x)\phi(P^*,x)&=-\f{zH_{n}(z,x)}{F_n(z,x)}, \lb{ch3.19}\\
\phi(P,x)+\phi(P^*,x)&=-2\f{zG_{n}(z,x)}{F_n(z,x)}, \lb{ch3.20}\\
\phi(P,x)-\phi(P^*,x)&=\f{2y}{F_n(z,x)}, \lb{ch3.21}
\end{align}
while $\Psi$ fulfills
\begin{align}
&\Psi_x(P,x,x_0)=U(z,x)\Psi(P,x,x_0), \lb{ch3.22} \\
&-y \Psi(P,x,x_0)=zV_n(z,x)\Psi(P,x,x_0), \lb{ch3.23} \\
&\psi_1(P,x,x_0)=\bigg(\f{F_n(z,x)}{F_n(z,x_0)}\bigg)^{1/2}
\exp\bigg(-(y/z)\int_{x_0}^x dx'F_n(z,x')^{-1} \bigg), \lb{ch3.24} \\
&\psi_1(P,x,x_0)\psi_1(P^*,x,x_0)=\f{F_n(z,x)}{F_n(z,x_0)},\lb{ch3.25} \\
&\psi_2(P,x,x_0)\psi_2(P^*,x,x_0)=-\f{H_n(z,x)}{zF_n(z,x_0)},\lb{ch3.26}
\\
&\psi_1(P,x,x_0)\psi_2(P^*,x,x_0)+\psi_1(P^*,x,x_0)\psi_2(P,x,x_0)
=2\f{G_{n}(z,x)}{F_n(z,x_0)}, \lb{ch3.27} \\
&\psi_1(P,x,x_0)\psi_2(P^*,x,x_0)-\psi_1(P^*,x,x_0)\psi_2(P,x,x_0)
=\f{2y}{zF_n(z,x_0)}. \lb{ch3.28}
\end{align}
In addition, as long as the zeros of $F_n(\dott,x)$ are all simple for 
$x\in\Omega$, $\Omega\subseteq\bbR$ an open interval, $\Psi(\dott,x,x_0)$, 
$x,x_0\in\Omega$, is meromorphic on $\calK_n\backslash\{P_0\}$. 
\end{lemma}

Next, we recall the Dubrovin-type equations for $\mu_j$. Since in the
remainder of this section we will frequently assume $\calK_n$ to be
nonsingular, we list all restrictions on $\calK_n$ in this case, 
\begin{equation}
E_m\in\bbC\backslash\{0\}, \; E_m\neq E_{m'} \text{ for } 
m\neq m', \, m,m'=0,\dots,2n+1, \; E_{2n+1}=0. \lb{ch3.30}
\end{equation}

\begin{lemma} \lb{lemma-ch3.2}  
Suppose $\eqref{ch3.0}$ and the $n$th stationary CH equation
$\eqref{ch2.29}$ holds subject to the constraint $\eqref{ch3.30}$ on an open
interval $\wti\Omega_\mu\subseteq\bbR$. Moreover,  suppose that the zeros
$\mu_j$, $j=1,\dots,n$, of $F_n(\dott)$ remain distinct and nonzero on
$\wti\Omega_\mu$. Then $\{\hat\mu_j\}_{j=1,\dots,n}$, defined by
\eqref{ch3.6},  satisfies the following first-order system of differential
equations
\begin{equation}
\mu_{j,x}(x)=2\f{y(\hat\mu_j(x))}{\mu_{j}(x)}
\prod_{\substack{\ell=1\\ \ell\neq j}}^n(\mu_j(x)-\mu_\ell(x))^{-1},
\quad j=1, \dots, n, \, x\in \wti\Omega_\mu. \lb{ch3.31}
\end{equation}
Next, assume $\calK_n$ to be nonsingular and introduce the initial
condition
\begin{equation}
\{\hat\mu_j(x_0)\}_{j=1,\dots,n}\subset\calK_n \lb{ch3.32}
\end{equation}
for some $x_0\in\bbR$, where $\mu_j(x_0)\neq 0$, $j=1,\dots,n$, are
assumed to be distinct. Then there exists an open interval
$\Omega_\mu\subseteq\bbR$, with $x_0\in\Omega_\mu$, such that the initial
value problem \eqref{ch3.31}, \eqref{ch3.32} has a unique solution
$\{\hat\mu_j\}_{j=1,\dots,n}\subset\calK_n$ satisfying
\begin{equation}
\hat\mu_j\in C^\infty(\Omega_\mu,\calK_n),\quad j=1, \dots, n,
\lb{ch3.33}
\end{equation}
and $\mu_j$, $j=1,\dots,n$, remain distinct and nonzero on $\Omega_\mu$.
\end{lemma}

Combining the polynomial approach in Section \ref{s2} with
\eqref{ch3.5} yields trace formulas for the CH invariants.
For simplicity we just record two simple cases.

\begin{lemma} \lb{lemma-ch3.3}
Suppose \eqref{ch3.0}, assume the $n$th stationary CH
equation \eqref{ch2.29} holds, and let $x\in\bbR$. Then 
\begin{align}
u(x)&=\f12\sum_{j=1}^n\mu_{j}(x)-\f14\sum_{m=0}^{2n+1} E_{m}, 
\lb{ch3.36} \\
4u(x)-u_{xx}(x)&=-\Bigg(\prod_{\substack{m=0\\E_m\neq 0}}^{2n+1}
E_m\Bigg)\Bigg( \prod_{j=1}^n \mu_j(x)^{-2}\Bigg). \lb{ch3.36a}
\end{align}
\end{lemma}

Next we turn to asymptotic properties of $\phi$ and $\psi_j$, $j=1,2$.

\begin{lemma} \lb{lemma-ch3.4}
Suppose \eqref{ch3.0}, assume the $n$th stationary CH
equation \eqref{ch2.29} holds, and let
$P=(z,y)\in\calK_n\backslash\{\Pinfp, \Pinfm,P_0\}$, $x\in\bbR$. Then
\begin{align}
&\phi(P,x)\underset{\zeta\to 0}{=}\begin{cases} 
-2\zeta^{-1}-2u(x)+u_x(x)+\Oh(\zeta), & P\to\Pinfp, \\
2u(x)+u_x(x)+\Oh(\zeta), & P\to\Pinfm, \end{cases} 
\quad \zeta=z^{-1},  \lb{ch3.38a} \\
&\phi(P,x)\underset{\zeta\to 0}{=}\Bigg(\prod_{\substack{m=0\\E_m\neq
0}}^{2n+1} E_m\Bigg)^{1/2}f_n(x)^{-1}\zeta +\Oh(\zeta^2),
\quad  P\to P_0, \quad \zeta=z^{1/2}, \lb{ch3.38b} 
\end{align}
and
\begin{align}
&\psi_1(P,x,x_0)\underset{\zeta\to 0}{=}\exp(\pm(x-x_0))(1+\Oh(\zeta)),
\quad P\to\Pinfpm, \quad \zeta=1/z, \lb{ch3.38c} \\
&\psi_2(P,x,x_0)\underset{\zeta\to 0}{=}\exp(\pm(x-x_0))
 \begin{cases}  -2+\Oh(\zeta), & P\to\Pinfp, \\
(2u(x)+u_x(x))\zeta +\Oh(\zeta^2), & P\to\Pinfm, \end{cases}
\lb{ch3.38d} \\ 
& \hspace*{9.96cm} \zeta=1/z, \no \\
& \psi_1(P,x,x_0)\underset{\zeta\to
0}{=}\exp\Bigg(-\f{1}{\zeta}\int_{x_0}^x dx'
\Bigg(\prod_{\substack{m=0\\E_m\neq 0}}^{2n+1} E_m
\Bigg)^{1/2}f_n(x')^{-1}+\Oh(1)\Bigg), \lb{ch3.38e}  \\ 
& \hspace*{7.6cm} P\to P_0, \; \zeta=z^{1/2}, \no\\
& \psi_2(P,x,x_0)\underset{\zeta\to
0}{=}\Oh\big(\zeta^{-1}\big)\exp\Bigg(-\f{1}{\zeta}\int_{x_0}^x dx'
\Bigg(\prod_{\substack{m=0\\E_m\neq 0}}^{2n+1} E_m
\Bigg)^{1/2}f_n(x')^{-1}+\Oh(1)\Bigg),  \no \\
& \hspace*{8cm} P\to P_0, \; \zeta=z^{1/2}. \lb{ch3.38f} 
\end{align} 
\end{lemma}

Since the representations of $\phi$ and $u$ in
terms of the Riemann theta function associated with $\calK_n$ (assuming
$\calK_n$ to be nonsingular) are not explicitly needed in this paper, we
omit the corresponding details and refer to the detailed treatment in 
\cite[Ch.\ 5]{GesztesyHolden:2000} and \cite{GH02} instead. 

Finally, we will recall that solvability of the Dubrovin equations 
\eqref{ch3.31} on $\Omega_\mu\subseteq\bbR$ in fact implies equation
\eqref{ch2.29}  on $\Omega_\mu$. 

\begin{theorem}\lb{theorem-ch3.10}
Fix $n\in\bbN$, assume \eqref{ch3.30}, and suppose that
$\{\hat\mu_j\}_{j=1,\dots,n}$ satisfies the stationary Dubrovin equations
\eqref{ch3.31} on an open interval $\Omega_\mu\subseteq\bbR$ such that
$\mu_j$, $j=1,\dots,n$, remain distinct and nonzero on $\Omega_\mu$. Then 
$u\in C^\infty(\Omega_\mu)$ defined by 
\begin{equation}
u(x)=\f12\sum_{j=1}^n\mu_j(x)-\f{1}{4}\sum_{m=0}^{2n+1} E_m \lb{ch3.40}
\end{equation}
satisfies the $n$th stationary CH equation \eqref{ch2.29}, that is,
\begin{equation}
\sCH_n(u)=0 \text{  on $\Omega_\mu$.} \lb{ch3.41}
\end{equation}
\end{theorem}

\section{Basic Facts on Self-adjoint Hamiltonian Systems} \lb{s4}

We now turn to the Weyl--Titchmarsh theory for singular Hamiltonian
(canonical) systems and briefly recall the basic material needed in the
following section. This material is standard and can be found, for
instance, in \cite{CG02}, \cite{HS81}, \cite{HS83}, \cite{HS84},
\cite{KR74}, \cite{LM02}, and the references therein.

\begin{hypothesis} \lb{h2.1}
$(i)$ Define the $2\times 2$ matrix 
$J=\left(\begin{smallmatrix}0& -1 \\ 1 & 0  \end{smallmatrix}\right)$,
and suppose $a_{j,k}, b_{j,k} \in L_{\loc}^1(\bbR)$, $j,k = 1,2$ and  
$A(x)=\big(a_{j,k}(x)\big)_{j,k=1,2}\ge 0$, 
$B(x)=\big(b_{j,k}(x)\big)_{j,k=1,2}=B(x)^*$ for a.e.~$x\in \bbR$.
We consider the Hamiltonian system 
\begin{equation}\lb{HSa}
J \varPsi'(z,x)=(zA(x)+B(x))\varPsi(z,x), \quad z\in\bbC
\end{equation}
for a.e. $x\in \bbR$, where $z$ plays the role of the spectral
parameter, and where
\begin{equation}\lb{HSb}
\varPsi(z,x) = (\psi_1(z,x)\;\psi_2(z,x))^\top, \quad
\psi_j(z,\dott)\in AC_{\loc}(\bbR), \,\, j=1,2.
\end{equation}
Here $AC_{\loc}(\bbR)$ denotes the set of locally absolutely
continuous functions on $\bbR$ and the $M^*$ and $M^\top$ denote the
adjoint and transpose of a matrix $M$, respectively. \\
$(ii)$ For all nontrivial solutions $\varPsi$ of \eqref{HSa} we assume
the definiteness hypothesis $($cf.\ \cite[Sect.\ 9.1]{At64}$)$
\begin{equation}\lb{2.3}
\int_{c}^d dx \, \varPsi(z,x)^*A(x)\varPsi(z,x) > 0\, ,
\end{equation}
on every interval $(c,d)\subset \bbR$, $c<d$. 
\end{hypothesis}
A simple example of a Hamiltonian system satisfying \eqref{2.3} is
obtained when
$A(x)= \left(\begin{smallmatrix} w(x) & 0 \\ 0 & 0 
\end{smallmatrix}\right)$, for some weight function 
$w\in L^1_{\loc}(\bbR)$, $w>0$ a.e. on $\bbR$, and $b_{2,2}(x)>0$ a.e. on
$\bbR$ (cf.\ Section \ref{s5}). Hypothesis \ref{h2.1}\,(ii) clearly holds
in this case.

Next, we introduce the vector space ($-\infty\leq a<b\leq\infty$)
\begin{equation}
L^2_A((a,b))=\bigg\{\phi: (a,b)\to\bbC^2 
\text{ measurable}\,\bigg|\, \int_a^b
dx\,(\phi(x),A(x)\phi(x))_{\bbC^2}<\infty \bigg\}, \lb{4.4}
\end{equation}
where $(\phi,\psi)_{\bbC^2}=\sum_{j=1}^2 \ol{\phi_j}\psi_j$ denotes the
standard scalar product in $\bbC^2$. Fix a point $x_0\in\bbR$. Then the
Hamiltonian system \eqref{HSa} is said to be in the {\it limit point case}
at $\infty$ (resp., $-\infty$) if for some (and hence for all)
$z\in\bbC\backslash\bbR$, precisely one solution of \eqref{HSa} lies in
$L^2_A((x_0,\infty))$ (resp., $L^2_A((-\infty,x_0))$). (By the analog
of Weyl's alternative, if \eqref{HSa} is not in the limit point case at
$\pm\infty$,  all solutions of \eqref{HSa} lie in $L^2_A((x_0,\pm\infty))$
for all $z\in\bbC$. In the latter case the Hamiltonian system
\eqref{HSa} is said to be in the {\it limit circle case} at $\pm\infty$.)

To simplify matters for the remainder of this section, we will
always suppose the limit point case at $\pm\infty$ from now on.
\begin{hypothesis} \lb{h2.2}
Assume Hypothesis \ref{h2.1} and suppose that the Hamiltonian system
\eqref{HSa} is in the limit point case at $\pm\infty$.
\end{hypothesis}
An elementary example of a Hamiltonian system satisfying Hypothesis
\ref{h2.2} is given by the case where all entries of $A$ and $B$ are
essentially bounded on $\bbR$ (cf.\ Section \ref{s5}).

When considering the Hamiltonian system \eqref{HSa} on the half-line
$[x_0,\infty)$ (resp., $(-\infty,x_0]$), a self-adjoint (separated)
boundary condition at the point $x_0$ is of the type
\begin{equation}
\alpha\Psi(x_0)=0, \lb{4.5}
\end{equation}
where $\alpha=(\alpha_1\;\alpha_2)\in\bbC^{1\times 2}$ satisfies
\begin{equation}
\alpha\alpha^*=I, \quad \alpha J\alpha^*=0 \; 
\text{ (equivalently, $|\alpha_1|^2+|\alpha_2|^2=1$, 
$\Im(\alpha_2 \ol{\alpha_1})=0$).} \lb{4.6}
\end{equation}
In particular, the boundary condition \eqref{4.5} (with $\alpha$
satisfying \eqref{4.6}) is equivalent to
$\alpha_1\psi_1(x_0)+\alpha_2\psi_2(x_0)=0$ with
$\alpha_1/\alpha_2\in\bbR$ if $\alpha_2\neq 0$ and 
$\alpha_2/\alpha_1\in\bbR$ if $\alpha_1\neq 0$. The special case
$\alpha_0=(1\; 0)$ will be of particular relevance in Section \ref{s5}. Due
to our limit point assumption at $\pm\infty$ in Hypothesis
\ref{h2.2}, no additional boundary condition at $\pm\infty$ needs to
be introduced when considering \eqref{HSa} on the half-lines
$[x_0,\infty)$ and $(-\infty,x_0]$. The resulting full-line and
half-line Hamiltonian systems are said to be self-adjoint on $\bbR$,
$[x_0,\infty)$, and $(-\infty,x_0]$, respectively (assuming of course a
boundary condition of the type \eqref{4.5} in the two half-line cases).

Next we digress a bit and briefly turn to Herglotz functions and their
representations in terms of measures, the focal point of
Weyl--Titchmarsh theory (and hence spectral theory) of self-adjoint
Hamiltonian systems.
\begin{definition} \lb{d4.9a}
Any analytic map $m\colon\bbC_+\to\bbC_+$ is called a {\it Herglotz}
function (here $\bbC_+=\{z\in\bbC\,|\,\Im(z)>0\}$). Similarly, any
analytic map $M\colon\bbC_+\to\bbC^{k\times k}$,
$k\in\bbN$, is called a $k\times k$ matrix-valued Herglotz function if 
$\Im(M(z))\geq 0$ for all $z\in\bbC_+$.
\end{definition}

Herglotz functions are characterized by a representation of the form 
\begin{align}
&m(z)= a + bz  +\int_{-\infty}^\infty
d\omega(\lambda)\,\big((\lambda-z)^{-1} -\lambda(1+\lambda^2)^{-1}\big),
\quad z\in\bbC\backslash\bbR, \lb{HF} \\ 
&a \in\bbR, \; b\geq 0, \quad \int_{-\infty}^{\infty} 
d\omega(\lambda)\,(1+\lambda^2)^{-1}  < \infty, \lb{HFa} \\
&\omega((\lambda, \mu])= \lim_{\delta\downarrow 0}
\lim_{\epsilon \downarrow 0}\frac{1}{\pi}\int_{\lambda
+ \delta}^{\mu + \delta }d\nu\,  \Im\left( m(\nu
+i\epsilon)\right), \lb{HFb}
\end{align}
in the following sense: Every Herglotz function admits a representation
of the type \eqref{HF}, \eqref{HFa} and conversely, any function of the
type \eqref{HF}, \eqref{HFa} is a Herglotz function. Moreover, local 
singularities and zeros of $m$ are necessarily located on the real axis and
at most of first order in the sense that
\begin{align}
&\omega(\{\lambda\})=\lim_{\varepsilon\downarrow0}\left(\omega(\lambda
+\varepsilon)-\omega(\lambda-\varepsilon)\right)=
-\lim_{\varepsilon\downarrow0} i\varepsilon\,
m(\lambda+i\varepsilon) \geq 0, \quad \lambda\in\bbR, \lb{4.10} \\
&\lim_{\varepsilon\downarrow0} i\varepsilon \, m(\lambda
+i\varepsilon)^{-1}\geq 0, \quad \lambda\in\bbR. \lb{4.11}
\end{align}
In particular, isolated poles of $m$ are simple and located
on the real axis, the corresponding residues being negative. Analogous
results hold for matrix-valued Herglotz functions (see, e.g., \cite{GT97}
and the literature cited therein).

For subsequent purpose in Section \ref{s5} we also note that
$-1/z$ is a Herglotz function and compositions of Herglotz functions remain
Herglotz functions. In addition, diagonal elements of a matrix-valued
Herglotz function are Herglotz functions. 

Returning to Hamiltonian systems on the half-lines $[x_0,\pm\infty)$
satisfying Hypotheses \ref{h2.1} and \ref{h2.2}, we now denote by
$\Psi_\pm(z,x,x_0)$ the unique solution of \eqref{HSa} satisfying
$\Psi_\pm(z,\dott,x_0)\in L^2_A([x_0,\pm\infty))$,
$z\in\bbC\backslash\bbR$, normalized by $\psi_{1,\pm}(z,x_0,x_0)=1$. Then
the half-line Weyl--Titchmarsh function $m_\pm(z,x)$, associated with the
Hamiltonian system \eqref{HSa} on $[x,\pm\infty)$ and the fixed boundary
condition $\alpha_0=(1\;0)$ at the point $x\in\bbR$, is defined by
\begin{equation}
m_\pm(z,x)=\psi_{2,\pm}(z,x,x_0)/\psi_{1,\pm}(z,x,x_0), \quad
z\in\bbC\backslash\bbR, \; \pm x\geq x_0.  \lb{4.12}
\end{equation}
The actual normalization of $\Psi_\pm(z,x,x_0)$ was chosen for
convenience only and is clearly immaterial in the definition of
$m_\pm(z,x)$ in \eqref{4.12}. 

One easily verifies that
$m_\pm(z,x)$ satisfies the following Riccati-type differential equation, 
\begin{align}
&m'(z,x) +[b_{2,2}(x)+a_{2,2}(x)z]m(z,x)^2 \lb{4.13} \\
&\quad + [b_{1,2}(x)+b_{2,1}(x)+(a_{1,2}(x)+a_{2,1}(x))z]m(z,x)  
+ b_{1,1}(x) +a_{1,1}(x)z =0. \no 
\end{align}

Finally, the $2\times 2$ Weyl--Titchmarsh matrix $M(z,x)$ associated
with the Hamiltonian system  \eqref{HSa} on $\bbR$ is then defined in
terms of the half-line Weyl--Titchmarsh functions $m_\pm(z,x)$ by
\begin{align}
M(z,x)
&=\big(M_{j,j^\prime}(z,x)\big)_{j,j^\prime=1,2},  \quad 
z\in\bbC\backslash\bbR,  \lb{4.14} \\
M_{1,1}(z,x)&=[m_-(z,x)-m_+(z,x)]^{-1}, \no \\ 
M_{1,2}(z,x)&=M_{2,1}(z,x) \no  \\
&=2^{-1}[m_-(z,x)-m_+(z,x)]^{-1}
[m_-(z,x)+m_+(z,x)], \lb{4.15} \\
M_{2,2}(z,x)&=[m_-(z,x)-m_+(z,x)]^{-1}
m_-(z,x)m_+(z,x). \no 
\end{align}
One verifies that $M(z,x)$ is a $2\times 2$ matrix-valued Herglotz
function. We emphasize that for any fixed $x_0\in\bbR$, $M(z,x_0)$
contains all the spectral information of the self-adjoint Hamiltonian
system \eqref{HSa} on $\bbR$ (assuming Hypotheses \ref{h2.1} and
\ref{h2.2}).

\section{Real-Valued Algebro-geometric CH Solutions \\ and the Associated
Isospectral Torus} \lb{s5}

In our final and principal section we study real-valued algebro-geometric
solutions of the CH hierarchy associated with curves $\calK_n$ whose
affine part is nonsingular and determine the isospectral manifold of
smooth bounded CH solutions. We focus on the stationary case  as
this is the primary concern in this context.  

To study the direct spectral problem we first introduce the following
assumptions.
\begin{hypothesis} \lb{h5.1}
Suppose 
\begin{equation}
E_0<E_1<\cdots <E_{2n}<E_{2n+1}=0 \lb{5.1}
\end{equation}
and let $u$ be a real-valued solution of the $n$th stationary CH equation
\eqref{ch2.29},
\begin{equation}
\sCH_{n}(u)=0, \lb{5.2}
\end{equation}
$($i.e., $u$ is a particular algebro-geometric CH potential\,$)$,
satisfying 
\begin{align}
&u\in C^\infty(\bbR), \quad \partial_x^k u \in L^\infty(\bbR), \;
k=0,1,2, \lb{5.3} \\ 
& 4u-u_{xx} > 0. \lb{5.4} 
\end{align}
\end{hypothesis}

We start by noticing that the basic stationary equation \eqref{ch3.22},
\begin{equation}
\Psi_x(z,x)=U(z,x)\Psi(z,x), \quad \Psi=(\psi_1, \psi_2)^\top, \;
(z,x)\in\bbC\times\bbR, \lb{5.5} 
\end{equation}
is equivalent to the following Hamiltonian (canonical) system
\begin{equation} 
J\wti\Psi_x(\ti z,x)=[\ti z A(x)+B(x)]\wti \Psi(\ti z,x), \quad 
\wti\Psi=(\ti\psi_1,\ti\psi_2)^\top, \;
(\ti z,x)\in\bbC\times\bbR, \lb{5.6} 
\end{equation}
where
\begin{align}
& J=\begin{pmatrix} 0 & -1 \\ 1 & 0 \end{pmatrix}, \quad 
\ti\Psi(\ti z,x)=\Psi(z,x), \quad \ti z=-1/z, \lb{5.7} \\
&A(x)=\begin{pmatrix} 4u(x)-u_{xx}(x) & 0 \\ 0 & 0 \end{pmatrix}, \quad 
B(x)=\begin{pmatrix} 0 & -1 \\ -1 & 1 \end{pmatrix}, \quad x\in\bbR.
\lb{5.8} 
\end{align}
In particular, due to assumptions \eqref{5.3} and \eqref{5.4}, the
Hamiltonian system \eqref{5.6} satisfies Hypotheses \ref{h2.1} and
\ref{h2.2}. Explicitly, the Hamiltonian system
\eqref{5.6} boils down to
\begin{align}
\ti\psi_{1,x}(\ti z,x)&=\ti\psi_2(\ti z,x)-\ti\psi_1(\ti z,x), \lb{5.8a} \\
\ti\psi_{2,x}(\ti z,x)&=-\ti z(4u(x)-u_{xx}(x))\ti\psi_1(\ti z,
x)+\ti\psi_2(\ti z,x), \quad (z,x)\in\bbC\times\bbR, \lb{5.8b}
\end{align}
and upon eliminating $\ti\psi_2$ results in a particular case of the
weighted Sturm--Liouville problem
\begin{equation}
\f{1}{r}\bigg[-\f{d}{dx}p\f{d}{dx}+q\bigg] \lb{5.8c}
\end{equation}
of the type
\begin{equation}
-\ti\psi_{1,xx}(\ti z,x)+\ti\psi_1(\ti z,x)=\ti
z(4u(x)-u_{xx}(x))\ti\psi_1(\ti z,x), \quad
(z,x)\in\bbC\times\bbR, \lb{5.8d}
\end{equation}
with ``weight'' $r=(4u-u_{xx})$ and constant coefficients $p=q=1$.

Introducing 
\begin{equation}
\Sigma=\bigcup_{\ell=0}^{n} [E_{2\ell},E_{2\ell+1}], \lb{5.9}
\end{equation} 
we define
\begin{align}
&R_{2n+2} (\lambda)^{1/2} = |R_{2n+2} (\lambda)^{1/2}| \no \\
&\times \begin{cases}
-1
& \text{for $\lambda \in (E_{2n+1}, \infty)$}, \\
(-1)^{n+j}
& \text{for $\lambda \in (E_{2j-1}, E_{2j})$,
$j=1,\dots,n$},  \\
(-1)^n
& \text{for $\lambda \in (-\infty, E_0)$},  \\
i(-1)^{n+j+1}  
& \text{for $\lambda \in (E_{2j}, E_{2j+1})$, 
$j=0,\dots,n$}, \end{cases} \qquad \lambda\in\bbR, \lb{5.11} 
\end{align}
and
\begin{equation}
{R}_{2n+2} (\lambda)^{1/2}
=\lim\limits_{\varepsilon \downarrow 0} {R}_{2n+2} 
(\lambda +i\varepsilon)^{1/2}, \quad \lambda \in \Sigma,
\lb{5.10}
\end{equation}
and analytically continue $R_{2n+2} (\dott)^{1/2}$ to
$\bbC\backslash\Sigma$. We also note the property
\begin{equation}
\ol{R_{2n+2}(\ol z)^{1/2}}=R_{2n+2}(z)^{1/2}. \lb{5.11a}
\end{equation}
For notational convenience we will occasionally call $(E_{2j-1},E_{2j})$,
$j=1,\dots,n$, spectral gaps and $E_{2j-1}, E_{2j}$ the corresponding
spectral gap endpoints.

Next, we introduce the cut plane
\begin{equation}
\Pi=\bbC\backslash\Sigma, \lb{5.12}
\end{equation}
and the upper, respectively, lower sheets $\Pi_\pm$ of $\calK_n$ by
\begin{equation}
\Pi_\pm = \{ (z, \pm R_{2n+2} (z)^{1/2})
\in \calK_n\mid z\in\Pi\} \lb{5.13}
\end{equation}
with the associated charts
\begin{equation}
\zeta_\pm \colon \Pi_\pm \to \Pi,\quad P=(z,\pm R_{2n+2}(z)^{1/2}) 
\mapsto z. \lb{5.14}
\end{equation}
The two branches $\Psi_{\pm}(z,x,x_0)$ of the
Baker--Akhiezer vector $\Psi(P,x,x_0)$ in \eqref{ch3.15} are then given by
\begin{equation}
\Psi_\pm (z,x,x_0)=\Psi(P,x,x_0), \quad P=(z,y)\in\Pi_\pm, \quad 
\Psi_\pm =(\psi_{1,\pm},\psi_{2,\pm})^\top, \lb{5.15}
\end{equation}
and one infers from \eqref{ch3.38c} that 
\begin{equation}
\psi_{1,\pm} (z,\dott,x_0)\in L^2((x_0,\mp\infty)) \, \text{ for $|z|$
sufficiently large.} \lb{5.16} 
\end{equation}
Thus, introducing 
\begin{equation}
\wti\Psi_\pm (\ti z,x,x_0)=\Psi_\mp(z,x,x_0), \quad 
\wti\Psi_\pm =(\ti\psi_{1,\pm},\ti\psi_{2,\pm})^\top, \quad \ti
z=-1/z, \lb{5.17}
\end{equation}
and the two branches $\phi_\pm(z,x)$ of $\phi(P,x)$ on $\Pi_\pm$ by
\begin{equation}
\phi_\pm (z,x)=\phi(P,x), \quad P=(z,y)\in\Pi_\pm, \lb{5.18}
\end{equation}
one infers from \eqref{4.12} and
\eqref{5.16} that the Weyl--Titchmarsh functions $\ti m_\pm(\ti z,x)$
associated with the self-adjoint Hamiltonian system
\eqref{5.6} on the half-lines $[x,\pm\infty)$ and the Dirichlet boundary
condition indexed by $\alpha_0=(1\;0)$ at the point $x\in\bbR$, are given
by
\begin{align}
\ti m_\pm(\ti z,x)&=\ti\psi_{2,\pm}(\ti z,x,x_0)
/\ti\psi_{1,\pm}(\ti z,x,x_0)=
\psi_{2,\mp}(z,x,x_0)/\psi_{1,\mp}(z,x,x_0) \no \\
&=(-1/z)\phi_{\mp}(z,x), \quad z\in\bbC\backslash\Sigma. \lb{5.19} 
\end{align}
More precisely, \eqref{5.16} yields \eqref{5.19} only for $|z|$
sufficiently large. However, since by general principles 
$\ti m_\pm(\dott,x)$ are analytic in $\bbC\backslash\bbR$, and by
\eqref{ch3.11}, $\phi_\pm(\dott,x)$ are analytic in $\bbC\backslash\Sigma$,
one infers \eqref{5.19} by analytic continuation. In particular,
\eqref{5.16} extends to all $z\in\bbC\backslash\Sigma$, that is,
\begin{equation}
\psi_{1,\pm} (z,\dott,x_0)\in L^2((x_0,\mp\infty)), \quad 
z\in\bbC\backslash\Sigma. \lb{5.19a} 
\end{equation}

Next, we mention a useful fact concerning a special class of Herglotz
functions closely related to the problem at hand. The result must be
well-known to experts, but since we could not quickly locate a proof in the
literature, we provide the simple contour integration argument below.

\begin{lemma} \lb{l5.3}
Let $P_N$ be a monic polynomial of degree $N$. Then 
$P_N/R_{2n+2}^{1/2}$ is a Herglotz function if and only if one of
the following alternatives applies: \\
$(i)$ $N=n$ and
\begin{equation}
P_n(z)=\prod_{j=1}^n (z-a_j), \quad a_j\in [E_{2j-1},E_{2j}], \;
j=1,\dots,n. \lb{5.22}
\end{equation}
If \eqref{5.22} is satisfied, then $P_n/R_{2n+2}^{1/2}$ admits the
Herglotz representation
\begin{equation}
\f{P_n(z)}{R_{2n+2}(z)^{1/2}}=\f{1}{\pi}\int_\Sigma 
\f{|P_n(\lambda)|\,d\lambda}{|R_{2n+2}(\lambda)^{1/2}|}\f{1}{\lambda-z}, 
\quad z\in\bbC\backslash\Sigma. \lb{5.23}
\end{equation}
$(ii)$ $N=n+1$ and 
\begin{equation}
P_{n+1}(z)=\prod_{\ell=0}^n (z-b_\ell), \quad 
b_0 \in (-\infty, E_0], \;\, b_j\in [E_{2j-1},E_{2j}], \;
j=1,\dots,n. \lb{5.24}
\end{equation}
If \eqref{5.24} is satisfied, then $P_{n+1}/R_{2n+2}^{1/2}$ admits
the Herglotz representation
\begin{align}
&\f{P_{n+1}(z)}{R_{2n+2}(z)^{1/2}}=
\Re\bigg(\f{P_{n+1}(i)}{R_{2n+2}(i)^{1/2}}\bigg)+\f{1}{\pi}\int_\Sigma 
\f{|P_{n+1}(\lambda)|\,d\lambda}{|R_{2n+2}(\lambda)^{1/2}|}
\bigg(\f{1}{\lambda-z}-\f{\lambda}{1+\lambda^2}\bigg), \no \\
&\hspace*{9.5cm} z\in\bbC\backslash\Sigma. \lb{5.25} 
\end{align}
\end{lemma}
\begin{proof}
Since Herglotz functions are $\Oh(z)$ as $|z|\to\infty$ and
cannot vanish faster than $\Oh(1/z)$ as $|z|\to\infty$, we can confine
ourselves to the range $N\in\{n,n+1,n+2\}$. We start with the case $N=n$
and employ the following contour integration approach. Consider a closed 
oriented contour $\Gamma_{R,\varepsilon}$ which consists of the
clockwise oriented semicircle
$C_{\varepsilon}=\{z\in\bbC\,|\, z=E_0-\varepsilon \exp(-i\alpha), \,
-\pi/2\leq\alpha\leq \pi/2\}$ centered at $E_0$, the straight line
$L_+=\{z\in\bbC_+\,|\, z=E_0+x+i\varepsilon, \, 0\leq x \leq R\}$
(oriented from left to right), the following part of the counterclockwise
oriented circle of radius $(R^2+\varepsilon^2)^{1/2}$ centered at $E_0$,
$C_R=\{z\in\bbC\,|\, z=E_0+(R^2+\varepsilon^2)^{1/2}\exp(i\beta), \,
\arctan(\varepsilon/R) \leq \beta \leq 2\pi-\arctan(\varepsilon/R)\}$,
and the straight line $L_-=\{z\in\bbC_-\,|\, z=E_0+x-i\varepsilon, \,
0\leq x\leq R\}$ (oriented from right to left). Then, for
$\varepsilon>0$ small enough and
$R>0$ sufficiently large, one infers
\begin{align}
\f{P_n(z)}{R_{2n+2}(z)^{1/2}}&=\f{1}{2\pi
i}\oint_{\Gamma_{R,\varepsilon}}
\f{1}{\zeta-z}\f{P_n(\zeta)}{R_{2n+2}(\zeta)^{1/2}}d\zeta \no \\
&\hspace*{-.45cm} \underset{\varepsilon\downarrow 0, R\uparrow\infty}{=}
\f{1}{\pi} \int_{\Sigma} 
\f{1}{\lambda-z}\f{P_n(\lambda)d\lambda}{iR_{2n+2}(\lambda)^{1/2}}.
\lb{5.27}
\end{align} 
Here we used \eqref{5.11} to compute the contributions of the contour
integral along $[E_0,R]$ in the limit $\varepsilon\downarrow 0$ and note
that the integral over $C_R$ tends to zero as $R\uparrow \infty$ since 
\begin{equation}
\f{P_n(\zeta)}{R_{2n+2}(\zeta)^{1/2}}\underset{\zeta\to\infty}{=}
\Oh\big(|\zeta|^{-1}\big). \lb{5.28}
\end{equation}
Next, utilizing the fact that $P_n$ is monic and using \eqref{5.11} again,
one infers that $F_n(\lambda)d\lambda/[iR_{2n+2}(\lambda)^{1/2}]$
represents a positive measure supported on $\Sigma$ if and only if $P_n$
has precisely one zero in each of the intervals $[E_{2j-1},E_{2j}]$, 
$j=1,\dots,n$. In other words, 
\begin{equation}
\f{P_n(\lambda)}{iR_{2n+1}(\lambda)^{1/2}}=
\f{|P_n(\lambda)|}{|R_{2n+1}(\lambda)^{1/2}|}\geq 0 \,\text{ on } \Sigma 
\lb{5.29}
\end{equation}
if and only if $P_n$ has precisely one zero in each of the intervals 
$[E_{2j-1},E_{2j}]$, $j=1,\dots,n$. The Herglotz representation
\eqref{HF}, \eqref{HFa} then finishes the proof of \eqref{5.23}. 

In the case where $N=n+1$, the proof of \eqref{5.24} follows along similar 
lines taking into account the additional residues at $\pm i$ inside
$\Gamma_{R,\varepsilon}$ which are responsible for the real part on the
right-hand side of \eqref{5.25}.

Finally, in the case $N=n+2$, assume that $P_{n+2}/R_{2n+2}^{1/2}$ is a
Herglotz function. Then necessarily,
\begin{equation}
\f{P_{n+2}(z)}{R_{2n+2}(z)^{1/2}}=a+bz+\int_{E_0}^0 d\omega(\lambda)\, 
(\lambda-z)^{-1}, \quad 
z\in\bbC\backslash\Sigma \lb{5.30}
\end{equation}
for some $a\in\bbR$, $b\geq 0$, and some finite (positive) measure
$\omega$ supported on $[E_0,0]$, since
\begin{equation}
\lim_{\varepsilon\downarrow 0}
\Im(P_{n+2}(\lambda)R_{2n+2}(\lambda+i\varepsilon)^{-1/2})=0 \, 
\text{ for $\lambda>E_{2n+2}=0$ and $\lambda<E_0$.} \lb{5.31}
\end{equation}
In particular, \eqref{5.30} implies
\begin{equation}
P_{n+2}(z)R_{2n+2}(z)^{-1/2}\underset{|z|\to\infty}{=}bz+\Oh(1), \quad
b\geq 0. \lb{5.32}
\end{equation}
However, by \eqref{5.11}, one immediately infers
\begin{equation}
P_{n+2}(\lambda)R_{2n+2}(\lambda)^{-1/2}
\underset{\lambda\uparrow\infty}{=}-\lambda+\Oh(1). \lb{5.33}
\end{equation}
This contradiction dispenses with the case $N=n+2$.
\end{proof}

Now we are in position to state the following result concerning the
half-line and full-line Weyl--Titchmarsh functions associated with the
self-adjoint Hamiltonian system \eqref{5.6}. We denote by $\ti m_\pm(\ti
z,x)$ the Weyl--Titchmarsh $m$-functions corresponding to \eqref{5.6}
associated with the half-lines $(x,\pm\infty)$ and the Dirichlet boundary
condition indexed by $\alpha_0=(1\;0)$ at the point $x\in\bbR$, and by
$\wti M(\ti z,x)$ the $2\times 2$ Weyl--Titchmarsh matrix corresponding to
\eqref{5.6} on $\bbR$ (cf.\ \eqref{4.12}, \eqref{4.14}, and \eqref{4.15}).
Moreover, $\Sigma^o$ denotes the open interior of $\Sigma$ and the real
part of a matrix $M$ is defined as usual by $\Re(M)=(M+M^*)/2$.

\begin{theorem} \lb{t5.2}
Assume Hypothesis \ref{h5.1} and let $(z,x)\in\bbR\times
(\bbC\backslash\Sigma)$, $\ti z=-1/z$. Then
\begin{align}
\ti m_\pm(\ti z,x)&= \f{\pm R_{2n+2}(z)^{1/2}+zG_n(z,x)}{zF_n(z,x)}
\lb{5.20} \\
&=1\pm \Re\bigg(\f{R_{2n+2}(i)^{1/2}}{iF_n(i,x)}\bigg)+
\sum_{j=1}^n \f{G_n(\mu_j(x),x)(1\mp \varepsilon_j (x))}
{dF_n(\mu_j(x),x)/dz}\f{1}{z-\mu_j(x)} \no \\
& \quad \pm \f{1}{\pi} \int_\Sigma
\f{|R_{2n+2}(\lambda)^{1/2}|\, d\lambda}{|\lambda F_n(\lambda,x)|}
\bigg(\f{1}{\lambda -z}-\f{\lambda}{1+\lambda^2}\bigg), \lb{5.20a} 
\end{align}
where $\varepsilon_j(x)\in\{1,-1\}$, $j=1,\dots,n$, is chosen such that 
\begin{equation}
\f{G_n(\mu_j(x),x)\varepsilon_j(x)}{dF_n(\mu_j(x),x)/dz}\geq 0, \quad
j=1,\dots,n. \lb{5.20b}
\end{equation}
Moreover,
\begin{align}
\wti M(\ti z,x)&= \f{-1}{2R_{2n+2}(z)^{1/2}} \begin{pmatrix}
-H_n(z,x) & zG_n(z,x) \\ zG_n(z,x) & zF_n(z,x) \end{pmatrix} \lb{5.21} \\
&= \Re(\wti M(i,x))+\int_\Sigma d\Omega(\lambda,x)\,
\bigg(\f{1}{\lambda-z}-\f{\lambda}{1+\lambda^2}\bigg), \lb{5.21A}
\end{align}
where
\begin{equation}
\Omega(\lambda,x)=\f{1}{2\pi i R_{2n+2}(\lambda)^{1/2}}\begin{pmatrix}
H_n(\lambda,x) & -\lambda G_n(\lambda,x) \\ 
-\lambda G_n(\lambda,x) & -\lambda F_n(\lambda,x) \end{pmatrix}, \quad
\lambda \in\Sigma^o. \lb{5.21B}
\end{equation}  
The essential spectrum of the half-line Hamiltonian systems
\eqref{5.6} on $[x,\pm\infty)$ $($with any self-adjoint boundary condition 
at $x$$)$ as well as the essential spectrum of the Hamiltonian system
\eqref{5.6} on $\bbR$ is purely absolutely continuous and given by
\begin{equation}
\bigcup_{\ell=0}^{n-1} \big[-E_{2l}^{-1},-E_{2\ell+1}^{-1}\big]\cup
\big[-E_{2n}^{-1},\infty\big).  \lb{5.21a}
\end{equation}
The spectral multiplicities are simple in the half-line cases and of
uniform multiplicity two in the full-line case.
\end{theorem}
\begin{proof}
Equation \eqref{5.20} follows from \eqref{ch3.11}, \eqref{5.11}, and
\eqref{5.19}. Equation \eqref{5.21} is then a consequence of
\eqref{ch3.19}--\eqref{ch3.21}, \eqref{4.14}, \eqref{4.15},
\eqref{5.19}, and \eqref{5.20}. Different self-adjoint boundary conditions
at the point $x$ lead to different half-line Hamiltonian systems whose
Weyl--Titchmarsh functions are related by a linear fractional
transformation (cf., e.g., \cite{CG02}), which leads to the invariance of
the essential spectrum with respect to the boundary condition at $x$. In
order to prove the Herglotz representation \eqref{5.20a} one can follow
the corresponding computation for Schr\"odinger operators with
algebro-geometric potentials in \cite[Sect.\ 8.1]{Le87}. For this purpose
one first notes that by \eqref{5.25} also
$R_{2n+2}(z)^{1/2}/[zF_n(z,x)]$ is a Herglotz function. A contour
integration as in the proof of Lemma
\ref{l5.3} then proves
\begin{align}
\f{R_{2n+2}(z)^{1/2}}{zF_n(z,x)} 
&=\Re\bigg(\f{R_{2n+2}(i)^{1/2}}{iF_n(i,x)}\bigg)+
\sum_{j=1}^n \f{|R_{2n+2}(\mu_j(x))^{1/2}|}
{\mu_j(x)|dF_n(\mu_j(x),x)/dz|}\f{1}{z-\mu_j(x)} \no \\
& \quad + \f{1}{\pi}\int_\Sigma
\f{|R_{2n+2}(\lambda)^{1/2}|\, d\lambda}{|\lambda F_n(\lambda,x)|}
\bigg(\f{1}{\lambda -z}-\f{\lambda}{1+\lambda^2}\bigg). \lb{5.21C} \\
&=\Re\bigg(\f{R_{2n+2}(i)^{1/2}}{iF_n(i,x)}\bigg) - 
\sum_{j=1}^n \f{G_{n}(\mu_j(x),x)\varepsilon_j(x)}
{dF_n(\mu_j(x),x)/dz}\f{1}{z-\mu_j(x)} \no \\
& \quad + \f{1}{\pi}\int_\Sigma
\f{|R_{2n+2}(\lambda)^{1/2}|\, d\lambda}{|\lambda F_n(\lambda,x)|}
\bigg(\f{1}{\lambda -z}-\f{\lambda}{1+\lambda^2}\bigg). \lb{5.21D}
\end{align}
The only difference compared to the corresponding argument in the proof of
Lemma \ref{l5.3} concerns additional (approximate) semicircles of
radius $\varepsilon$ centered at each $\mu_j(x)$, $j=1,\dots,n$, in the
upper and lower complex half-planes. Whenever, $\mu_j(x)\in
(E_{2j-1},E_{2j})$, the limit $\varepsilon\downarrow 0$ picks up a residue
contribution displayed in the sum over $j$ in \eqref{5.21C}. This
contribution vanishes, however, if $\mu_j(x)\in\{E_{2j-1}, E_{2j}\}$. In
this case $dF_n(\mu_j(x),x)/dz\neq 0$ by \eqref{4.10} and
$R_{2n+2}(\mu_j(x))=0$ by \eqref{ch2.22}. Equation \eqref{5.21D} then
follows from \eqref{ch3.6} and the sign of $\varepsilon_j(x)$ must be
chosen according to \eqref{5.20b} in order to guarantee nonpositive
residues in \eqref{5.21D} (cf.\ \eqref{4.10}). 

Next, we apply the Lagrange interpolation formula. If $Q_{n-1}$ is a
polynomial of degree $n-1$, then
\begin{equation}
Q_{n-1}(z)=F_n(z)\sum_{j=1}^n
\f{Q_{n-1}(\mu_j)}{dF_n(\mu_j)/dz}\f{1}{z-\mu_j}, \quad z\in\bbC.
\lb{5.21E}
\end{equation}
Since $F_n$ and $G_n$ are monic polynomials of degree $n$, we can apply
\eqref{5.21E} to $Q_{n-1}=G_n-F_n$ and obtain
\begin{equation}
\f{G_n(z,x)}{F_n(z,x)}=1+\sum_{j=1}^n
\f{G_n(\mu_j(x),x)}{d F_n(\mu_j(x),x)/dz}\f{1}{z-\mu_j(x)}. \lb{5.21F}
\end{equation}
Insertion of \eqref{5.21F} into \eqref{5.20} then yields 
\begin{equation}
\tilde m_\pm(\tilde z,x)=\f{\pm R_{2n+2}(z)^{1/2}}{zF_n(z,x)}+
\sum_{j=1}^n \f{G_n(\mu_j(x),x)[\varepsilon_j(x)+(1-\varepsilon_j(x))]}
{d F_n(\mu_j(x),x)/dz} \f{1}{z-\mu_j(x)} +1, \lb{5.21G}
\end{equation}
and hence \eqref{5.20a} follows by inserting \eqref{5.21D} into
\eqref{5.21G}. Equations \eqref{5.21A} and \eqref{5.21B} are clear from
the matrix analog of \eqref{HFb}.

The statement \eqref{5.21a} for the essential half-line spectra then
follows from the fact that the measure in the Herglotz representation
\eqref{5.20a} of $\tilde m_\pm$ (as a function of $z$) is supported on the
set $\Sigma$ in \eqref{5.9}, with a strictly positive density on the open
interior $\Sigma^o$ of $\Sigma$. The transformation $z\to -1/z$ then
yields \eqref{5.21a} and since half-line spectra with a regular endpoint
$x$ have always simple spectra this completes the proof of our half-line
spectral assertions. The full-line case follows in exactly the same manner
since the corresponding $2\times 2$ matrix-valued measure $\Omega$ in the
Herglotz representation \eqref{5.21A} of $\wti M$ (as a function of
$z$) also has support $\Sigma$ and rank equal to two on $\Sigma^o$.
\end{proof}

Returning to direct spectral theory, we note that the two spectral  
problems \eqref{5.6} on $\bbR$ associated with the vanishing of the first
and second component of $\wti \Psi$ at $x$, respectively, are clearly
self-adjoint since they correspond to the choices $\alpha=(1\;0)$ and
$\alpha=(0\;1)$ in \eqref{4.5}. Hence a comparison with
\eqref{ch3.5},
\eqref{ch3.25}, and \eqref{ch3.26} necessarily yields 
$\{\mu_j(x)\}_{j=1,\dots,n}, \{\nu_j(x)\}_{j=1,\dots,n}\subset\bbR$. Thus
we will assume the  convenient eigenvalue orderings 
\begin{equation}
\mu_j(x) < \mu_{j+1}(x), \quad \nu_j(x) < \nu_{j+1}(x) \text{  for
$j=1,\dots,n-1$}, \; x\in\bbR.  \lb{5.34} 
\end{equation}
The zeros of $\ti\psi_1(\dott,x)$ belong to the Dirichlet spectral problem
associated with the Hamiltonian system \eqref{5.6} (resp., the weighted
Sturm--Liouville problem \eqref{5.8d}) on $\bbR$. A comparison with
\eqref{ch3.25} then relates the zeros $\mu_j(x_1)$, $j=1,\dots,n$, of
$F_n(\dott,x_1)$ in \eqref{ch3.5} to the Dirichlet spectrum of \eqref{5.6}
(resp., \eqref{5.8d}) on $\bbR$. The correspondence between each $\mu_j$
and the related spectral point of the Dirichlet problem \eqref{5.6} (resp.,
\eqref{5.8d}) on $\bbR$ is of course effected by the transformation
$z\to -1/z$. In contrast to this, the zeros of $\ti\psi_2(\dott,x_1)$ do
not belong to the Neumann spectrum associated with the Hamiltonian
system \eqref{5.6} (resp., the weighted Sturm--Liouville problem
\eqref{5.8d}) on $\bbR$. In fact, by \eqref{5.8a}, zeros of
$\ti\psi_2(\dott,x_1)$ correspond to a mixed boundary condition of the type
$\ti\psi_{1,x}(x_1)+\ti\psi_1(x_1)=0$. By \eqref{ch3.26}, this relates the 
zeros $\nu_j(x_1)$, $j=1,\dots,n$, of $H_n(\dott,x_1)$ in \eqref{ch3.5} to
the spectrum of \eqref{5.6} (resp., \eqref{5.8d}) on $\bbR$ corresponding
to the self-adjoint boundary condition
$\ti\psi_{1,x}(x_1)+\ti\psi_1(x_1)=0$.
 
Combining Lemma \ref{l5.3} with the Herglotz property of the 
$2\times 2$ Weyl--Titchmarsh matrix $\wti M(\dott,x)$ then yields the
following refinement of Theorem \ref{lemma-ch3.2}.

\begin{theorem} \lb{t5.4}
Assume Hypothesis 5.1. Then $\{\hat\mu_j\}_{j=1,\dots,n}$, with
the projections $\mu_j(x)$, $j=1,\dots,n$, the zeros of $F_n(\dott,x)$ in
\eqref{ch3.5}, satisfies the first-order system of differential equations
\eqref{ch3.31} on
$\Omega_\mu=\bbR$ and 
\begin{equation}
\hat\mu_j\in C^\infty(\bbR,\calK_n),\quad j=1, \dots, n. \lb{5.36}
\end{equation}
Moreover,
\begin{equation}
\mu_j(x)\in[E_{2j-1},E_{2j}], \quad j=1,\dots,n, \; x\in\bbR.
\lb{5.37}
\end{equation}
In particular, $\hat \mu_j (x)$ changes sheets whenever it hits 
$E_{2j-1}$ or $E_{2j}$ and its projection $\mu_j (x)$ remains trapped in
$[E_{2j-1}, E_{2j}]$ for all $j=1,\dots,n$ and $x\in\bbR$. The analogous
statements apply to $\hat \nu_j(x)$ and one infers
\begin{equation}
\nu_j(x)\in[E_{2j-1},E_{2j}], \quad j=1,\dots,n, \; x\in\bbR. \lb{5.38}
\end{equation}
\end{theorem}
\begin{proof}
Since $\wti M(\dott,x)$ is a $2\times 2$ Herglotz matrix, its diagonal
elements are Herglotz functions. Thus, 
\begin{equation}
\wti M_{1,1}(\ti z,x) =\f{H_n(z,x)}{2R_{2n+2}(z)^{1/2}}, \quad 
\wti M_{2,2}(\ti z,x) =\f{-zF_n(z,x)}{2R_{2n+2}(z)^{1/2}}, 
\quad \ti z=-1/z \lb{5.39}
\end{equation}
are Herglotz functions (the left-hand sides with respect to $\ti z$,
the right-hand sides with respect to $z$) and the interlacing properties
\eqref{5.37}, \eqref{5.38} then follow from \eqref{5.24} and \eqref{5.22}.
\end{proof}

\begin{remark} \lb{r5.5} 
Combining the interlacing property \eqref{5.37} with
\eqref{ch2.26}, \eqref{ch2.27}, and \eqref{ch2.27a} yields 
$($cf.\ also \eqref{ch3.36a}$)$
\begin{equation}
4u(x)-u_{xx}(x)=-\bigg(\prod_{m=0}^{2n} E_m\bigg) 
\bigg(\prod_{j=1}^n \mu_j(x)^{-2}\bigg)>0, \quad x\in\bbR, \lb{5.40}
\end{equation}
in accordance with \eqref{5.4}. Moreover, since
by \eqref{5.38} the $\nu_j(x)$ also remain trapped in the intervals
$[E_{2j-1},E_{2j}]$ for all $x\in\bbR$, none of the $\hat\nu_j$ can reach
$\Pinfm$ and hence $h_0=4u+2u_x\neq 0$ on $\bbR$ $($cf.\ the discussion
surrounding \eqref{ch3.13a}$)$. Actually,
\begin{equation}
h_0(x)>0, \quad x\in\bbR, \lb{5.41}
\end{equation}
since $H_n(\dott,x)/R_{2n+2}^{1/2}$ is a Herglotz function $($cf.\
\eqref{5.23}$)$. 
\end{remark}

\begin{remark} \lb{r5.6}
The zeros $\mu_j(x)\in (E_{2j-1},E_{2j})$, $j=1,\dots,n$ of
$F_n(\dott,x)$ which are related to eigenvalues of the Hamiltonian system
\eqref{5.6} on $\bbR$ associated with the boundary condition
$\ti\psi_1(x)=0$, in fact, are related to left and right half-line
eigenvalues of the corresponding Hamiltonian system restricted to the
half-lines $(-\infty,x]$ and $[x,\infty)$, respectively. Indeed, by
\eqref{5.17} and \eqref{5.19a}, depending on whether
$\hat\mu_j(x)\in\Pi_+$ or $\hat\mu_j(x)\in\Pi_-$, $\mu_j(x)$ is related to
a left or right half-line eigenvalue associated with the Dirichlet
boundary condition $\ti\psi_1(x)=0$. A careful investigation of the sign
of the right-hand sides of the Dubrovin equations \eqref{ch3.30}
$($combining \eqref{5.1}, \eqref{5.11}, and \eqref{5.13}$)$, then proves
that the $\mu_j(x)$ related to right $($resp., left\,$)$ half-line
eigenvalues of the Hamiltonian system \eqref{5.6} associated with the
boundary condition $\ti\psi_1(x)=0$, are strictly monotone increasing
$($resp., decreasing\,$)$ with respect to $x$, as long as the $\mu_j$ stay
away from the right $($resp., left\,$)$ endpoints of the corresponding 
spectral gaps $(E_{2j-1},E_{2j})$. Here we purposely avoided the limiting
case where some of the $\mu_k(x)$ hit the boundary of the spectral gaps, 
$\mu_k(x)\in\{E_{2k-1},E_{2k}\}$, since the half-line eigenvalue
interpretation is lost as there is no $L^2((x,\pm\infty))^2$
eigenfunction $\wti\Psi(x)$ satisfying $\ti\psi_1(x)=0$ in this case. In
fact, whenever an eigenvalue $\mu_k(x)$ hits a spectral gap endpoint, the
associated point $\hat\mu_j(x)$ on $\calK_n$ crosses over from one
sheet to the other $($equivalently, the corresponding left half-line
eigenvalue becomes a right half-line eigenvalue and vice versa\,$)$ and
accordingly, strictly increasing half-line eigenvalues become strictly
decreasing half-line eigenvalues and vice versa. In particular, using the
appropriate local coordinate $(z-E_{2k})^{1/2}$ $($resp.,
$(z-E_{2k-1})^{1/2}$$)$ near
$E_{2k}$ $($resp., $E_{2k-1}$$)$, one verifies that $\mu_k(x)$ does not
pause at the endpoints $E_{2k}$ and $E_{2k-1}$.
\end{remark}

Next, we turn to the inverse spectral problem and determine the
isospectral manifold of real-valued, smooth, and bounded CH solutions.

Our basic assumptions then will be the following:
\begin{hypothesis} \lb{h5.7}
Suppose 
\begin{equation}
E_0<E_1<\cdots <E_{2n}<E_{2n+1}=0, \lb{5.42}
\end{equation}
fix $x_0\in\bbR$, and assume that the initial data
\begin{equation}
\{\hat\mu_j(x_0)=(\mu_j(x_0),-\mu_j(x_0)
G_{n}(\mu_j(x_0),x_0))\}_{j=1,\dots,n} \subset\calK_n \lb{5.43}
\end{equation}
for the Dubrovin equations \eqref{ch3.31} are constrained by
\begin{equation}
\mu_j(x_0)\in[E_{2j-1},E_{2j}], \quad j=1,\dots,n. \lb{5.44}
\end{equation}
\end{hypothesis}

\begin{theorem} \lb{t5.8}
Assume Hypothesis \ref{h5.7}. Then the Dubrovin initial value problem
\eqref{ch3.31}, \eqref{5.43}, \eqref{5.44} has a unique solution
$\{\hat\mu_j\}_{j=1,\dots,n}\subset\calK_n$ satisfying
\begin{equation}
\hat\mu_j\in C^\infty(\bbR,\calK_n),\quad j=1, \dots, n,
\lb{5.45}
\end{equation}
and the projections $\mu_j$ remain trapped in the intervals
$[E_{2j-1},E_{2j}]$, $j=1,\dots,n$, for all $x\in\bbR$,
\begin{equation}
\mu_j(x)\in[E_{2j-1},E_{2j}], \quad j=1,\dots,n, \; x\in\bbR. \lb{5.46}
\end{equation}
Moreover, $u$ defined by the trace formula \eqref{ch3.36}, that is,
\begin{equation}
u(x)=\f12\sum_{j=1}^n\mu_{j}(x)-\f14\sum_{m=0}^{2n+1} E_{m}, 
\quad x\in\bbR, \lb{5.47}
\end{equation}
satisfies Hypothesis \ref{h5.1}, that is, 
\begin{align}
&u\in C^\infty(\bbR), \quad u \text{ is real-valued,} \lb{5.48} \\ 
&\partial_x^k u \in L^\infty(\bbR), \; k\in\bbN_0, \lb{5.49} \\ 
& 4u-u_{xx} > 0, \quad x\in\bbR, \lb{5.50} 
\intertext{and the $n$th stationary CH equation}
&\sCH_{n}(u)=0 \text{ on $\bbR$}, \lb{5.51}
\end{align}
with integration constants $c_\ell$ in \eqref{5.51} given by
$c_\ell=c_\ell(\ul E)$, $\ell=1,\dots,n$, according to \eqref{ch2.39C},
\eqref{ch2.39D}. 
\end{theorem}
\begin{proof}
Given initial data constrained by $\mu_j(x_0)\in(E_{2j-1},E_{2j})$, 
$j=1,\dots,n$, one concludes from the Dubrovin equations \eqref{ch3.31} and
the sign properties of $R_{2n+2}^{1/2}$ on the intervals
$[E_{2k-1},E_{2k}]$, $k=1,\dots,n$, described in \eqref{5.11}, that the
solution $\mu_j(x)$ remains in the interval $[E_{2j-1},E_{2j}]$ as long as
$\hat\mu_j(x)$ stays away from the branch points $(E_{2j-1},0),
(E_{2j},0)$. In case $\hat \mu_j$ hits  such a branch point, one can use
the local chart around $(E_m,0)$, with local coordinate $\zeta=\sigma
(z-E_m)^{1/2}$, $\sigma\in\{1,-1\}$, $m\in\{2j-1,2j\}$, to verify
\eqref{5.45} and \eqref{5.46}. Relations \eqref{5.47}--\eqref{5.49} are
then evident from \eqref{5.45}, \eqref{5.46}, and
\begin{equation}
|\partial_x^k \mu_j(x)|\leq C_k, \quad k\in\bbN_0, \; j=1,\dots,n, \;
x\in\bbR. \lb{5.52}
\end{equation}
In the course of the proof of Theorem \ref{theorem-ch3.10} presented in
\cite{GH02} (cf.\ also \cite[Sect.\ 5.3]{GesztesyHolden:2000}), one
constructs the polynomial formalism ($F_n$, $G_n$, $H_n$, $R_{2n+2}$, etc.)
and then obtains identity \eqref{ch3.36a} as an elementary consequence. The
latter immediately proves \eqref{5.50}. Finally, \eqref{5.51} follows from
Theorem \ref{theorem-ch3.10} (with $\Omega_\mu=\bbR$).
\end{proof}

\begin{corollary} \lb{c5.9}
Fix $\{E_m\}_{m=0,\dots, 2n+1}\subset\bbR$ and assume the ordering
\eqref{5.42}. Then the isospectral manifold of smooth bounded
real-valued solutions $u\in C^\infty(\bbR)\cap L^\infty(\bbR)$ of
$\sCH_{n}(u)=0$ is given by a real $n$-dimensional torus $\bbT^n$.
\end{corollary}
\begin{proof}
The discussion in Remark \ref{r5.6} and Theorem \ref{t5.8}, shows that the
motion of each $\hat\mu_j(x)$ on $\calK_n$ proceeds topologically on a
circle and is uniquely determined by the initial data $\hat\mu_k(x_0)$,
$k=1,\dots,n$. More precisely, the initial data 
\begin{align}
\begin{split}
&\hat\mu_j(x_0)=(\mu_j(x_0),y(\hat\mu_j(x_0)))=(\mu_j(x_0),
-\mu_j(x_0)G_n(\mu_j(x_0),x_0)), \lb{5.53} \\
&\mu_j(x_0)\in[E_{2j-1},E_{2j}], \quad j=1,\dots,n
\end{split}
\end{align}
are topologically equivalent to data of the type 
\begin{align}
(\mu_j(x_0),\sigma_j(x_0))\in [E_{2j-1},E_{2j}]\times \{+,-\},  
\quad j=1,\dots,n, \lb{5.54}
\end{align}
the sign of $\sigma_j(x_0)$ depending on $\hat\mu_j(x_0)\in\Pi_\pm$. If
some of the $\mu_k(x_0)\in\{E_{2k-1},E_{2k}\}$, then the determination of
the sheet $\Pi_\pm$ and hence the sign $\sigma_k(x_0)$ in \eqref{5.54}
becomes superfluous and is eliminated from \eqref{5.54}. Indeed, since by
\eqref{ch2.23}, 
\begin{equation}
\mu_j(x_0)^2G_n(\mu_j(x_0),x_0)=R_{2n+2}(\mu_j(x_0)), \lb{5.52a}
\end{equation}
$G_n(\mu_j(x_0),x_0)$ is determined up to a sign unless $\mu_j(x_0)$ hits
a spectral gap endpoint $E_{2j-1}, E_{2j}$ in which case
$G_n(\mu_j(x_0),x_0)=R_{2n+2}(\mu_j(x_0))=0$ and the sign ambiguity
disappears. The $n$ data in \eqref{5.54} (properly interpreted if
$\mu_j(x_0)\in\{E_{2j-1},E_{2j}\}$) can be identified with circles. Since
the latter are  independent of each other, the isospectral manifold of
real-valued, smooth, and bounded solutions of
$\sCH_{n}(u)=0$ is given by a real
$n$-dimensional torus $\bbT^n$.
\end{proof}

\begin{remark} \lb{r5.10} ${}$ \\
$(i)$ For simplicity we only focused on the case $4u-u_{xx}>0$. The
opposite case $4u-u_{xx}<0$ is completely analogous and results in a
reflection of $E_m$, $m=0,\dots,2n+1$, and $\mu_j(x),\nu_j(x)$,
$j=1,\dots,n$, about $z=0$, etc. \\
$(ii)$ The time-dependent case also offers nothing new. Higher-order
$\CH_r$ flows drive each $\hat\mu_j(x,t_r)$ around the same circles as in
the stationary case in complete analogy to the familiar KdV case.
\end{remark}

In summary, one observes that the reality problem for smooth bounded
solutions of the CH hierarchy, assuming the ordering \eqref{5.42} (resp.,
the one obtained upon reflection with respect to $z=0$), parallels that of
the KdV hierarchy with the basic self-adjoint Lax operator (the
one-dimensional Schr\"odinger operator) replaced by the self-adjoint
Hamiltonian system
\eqref{5.6}. 

The following result was found in response to a query of Igor Krichever,
who inquired about the significance of the eigenvalue ordering
\eqref{5.42} (or the one obtained upon reflection at $z=0$). As it turns
out, such an ordering is crucial if one is interested in smooth
algebro-geometric solutions on $\bbR$. 

\begin{theorem} \lb{t5.11}
Suppose 
\begin{align}
&E_m<E_{m+1}, \quad m=0,\dots 2n, \no \\ 
&\text{and $E_{2j_0-1}=0$ 
$($resp., $E_{2j_0}=0$$)$ for some $j_0\in\{1,\dots,n\}$.} \lb{5.61}
\end{align}
Fix $x_0\in\bbR$, and assume that the initial data
\begin{equation}
\{\hat\mu_j(x_0)=(\mu_j(x_0),-\mu_j(x_0)
G_{n}(\mu_j(x_0),x_0))\}_{j=1,\dots,n} \subset\calK_n \lb{5.62}
\end{equation}
for the Dubrovin equations \eqref{ch3.31} are constrained by
\begin{align}
&\mu_j(x_0)\in[E_{2j-1},E_{2j}], \quad j\in\{1,\dots,n\}\backslash\{j_0\}, 
\no \\
&\text{and $\mu_{j_0}(x_0)\in (E_{2j_0-1},E_{2j_0}]$ 
$($resp., $\mu_{j_0}(x_0)\in [E_{2j_0-1},E_{2j_0}))$.} \lb{5.63}
\end{align}
Then there exists a set $\Omega_\mu\subset\bbR$ of the type
\begin{equation}
\Omega_\mu= \bbR\backslash\{\xi_k\}_{k\in\bbZ}, \quad \xi_k<\xi_{k+1}, \;
k\in\bbZ, \quad \lim_{k\downarrow -\infty} \xi_k =-\infty, \;  
\lim_{k\uparrow \infty} \xi_k =\infty, \lb{5.64}
\end{equation}
such that the Dubrovin initial value problem \eqref{ch3.31}, \eqref{5.43},
\eqref{5.44} has a unique solution
$\{\hat\mu_j\}_{j=1,\dots,n}\subset\calK_n$ satisfying
\begin{equation}
\hat\mu_j\in C^\infty(\Omega_\mu,\calK_n),\quad j=1, \dots, n,
\lb{5.65}
\end{equation}
and the projections $\mu_j$ remain trapped in the intervals
$[E_{2j-1},E_{2j}]$, $j=1,\dots,n$, for all $x\in\Omega_\mu$,
\begin{equation}
\mu_j(x)\in[E_{2j-1},E_{2j}], \quad j=1,\dots,n, \; x\in\Omega_\mu.
\lb{5.66}
\end{equation}
Moreover, $u$ defined by the trace formula \eqref{ch3.36}, that is, 
\begin{equation}
u(x)=\f12\sum_{j=1}^n\mu_{j}(x)-\f14\sum_{m=0}^{2n+1} E_{m}, 
\quad x\in\Omega_\mu, \lb{5.67}
\end{equation}
satisfies 
\begin{align}
&u\in C^\infty(\Omega_\mu), \quad u \text{ is real-valued,} \lb{5.68} \\ 
& 4u-u_{xx} > 0, \quad x\in\Omega_\mu, \lb{5.69} 
\intertext{and the $n$th stationary CH equation}
&\sCH_{n}(u)=0 \text{ on $\Omega_\mu$}, \lb{5.70}
\end{align}
with integration constants $c_\ell$ in \eqref{5.51} given by
$c_\ell=c_\ell(\ul E)$, $\ell=1,\dots,n$, according to \eqref{ch2.39C},
\eqref{ch2.39D}. At each $\xi_k$, $u_x$ exhibits a singularity of the type 
\begin{equation}
u_x(x)\underset{x\to \xi_k}{=} C_k(x-\xi_k)^{-1/3} + 
\oh\big((x-\xi_k)^{-1/3}\big), \quad C_k\neq 0, \; k\in\bbZ. \lb{5.71}
\end{equation}
In particular, $u\notin C^1(\bbR)$, $u_x\notin L^\infty(\bbR)$. The
isospectral manifold corresponding to \eqref{5.61}--\eqref{5.63} is then
given by $\bbT^{n-1}\times\bbR$.
\end{theorem}
\begin{proof}
One can follow the proof of Theorem \ref{t5.8} and Corollary
\ref{c5.9} with one important twist, though, since the right-hand side of
the Dubrovin equation of $\mu_{j_0}$ blows up as $\mu_{j_0}\to E_{2j_0-1}
\text{\,(resp., $E_{2j_0}$)}$. For notational convenience and without loss
of generality we may assume $E_{1}=0$ (and hence $j_0=1$) in the following.
Recalling the Dubrovin equations \eqref{ch3.31}, one verifies that its
solutions are smooth with respect to $x$ as long as $\mu_1$ stays away
from $E_1=0$. Varying $x\in\bbR$, the sign restrictions on
$\mu_{1,x}$ in terms of the right-hand side of the corresponding equation
in \eqref{ch3.31} eventually accelerate $\mu_1$ into $E_1=0$ as $x$ tends
to some $\xi_{k}\in\bbR$, and we now analyze what happens to all $\mu_j$
for $x$ in a neighborhood of $\xi_{k}$. Recalling the local coordinate
$\sigma z^{1/2}$, $\sigma=\pm 1$, near $E_{1}=0$ and hence
introducing 
\begin{equation}
\zeta_{1}(x)=\sigma (-\mu_{1}(x))^{1/2} \text{ for $\mu_{1}(x)$
sufficiently close to $E_{1}=0$ as $x\to \xi_{k'}$,} \lb{5.72}
\end{equation}
and the corresponding point $\hat\mu_1=\widehat{-\zeta_1^2}\in\calK_n$, the
Dubrovin equation for $\mu_{j}$ becomes for $x$ near $\xi_{k}$,
\begin{align}
\zeta_{1,x}(x)&=\f{y(\widehat{-\zeta_{1}(x)^2})}{\zeta_{1}(x)^3}
\prod_{\ell=2}^n(-\zeta_{1}(x)^2
-\mu_\ell(x))^{-1}, \lb{5.73} \\
&\hspace*{-2.5mm} \underset{x\to
\xi_{k'}}{=}C_{1}\zeta_{1}(x)^{-2}(1+\oh(1)), \no \\
\mu_{j,x}(x)&=2\f{y(\hat\mu_j(x))}{\mu_{j}(x)}
\prod_{\substack{\ell=1\\ \ell\neq j}}^n(\mu_j(x)-\mu_\ell(x))^{-1},
\quad j=2, \dots, n. \lb{5.74}
\end{align}
for some constant $C_{1}\neq 0$. (Here we implicitly assume that no other
$\mu_j$, $j=2,\dots,n$ simultaneously hits $E_{2j-1}$ or $E_{2j}$ as
$x\to\xi_{k}$. Otherwise one simply resorts to the proper local coordinate
for such a $\mu_j$. We omit the details.) To treat the singularity of
$\zeta_{1,x}$ as $\zeta_1(x)\to 0$ for $x\to \xi_{k}$, we now resort to a
well-known trick described, for instance, in
\cite[Theorem\ 3.2.2]{Hi76} in the context of scalar first-order
differential equations. Instead of looking for solutions $\zeta_1$,
$\mu_j$ as functions of $x$, we now look for $x=x(\zeta_1)$,
$\ti\mu_j=\ti\mu_j(\zeta_1)$ as functions of $\zeta_1$, where we denote
$\ti\mu_j(\zeta_1)=\mu_j(x)$, $j=2,\dots,n$. Then \eqref{5.73},
\eqref{5.74} turn into
\begin{align}
x'(\zeta_1)&=\f{\zeta_{1}^3}{y(\widehat{-\zeta_{1}^2})}
\prod_{\ell=2}^n(-\zeta_{1}^2
-\ti\mu_\ell(\zeta_1)), \lb{5.75} \\
\ti\mu_{j,\zeta_1}(x)&=
2\f{y(\hat{{\ti\mu}}_j(\zeta_1))}{y(\widehat{-\zeta_{1}^2})}
\f{\zeta_{1}^3}{\ti\mu_{j}(\zeta_1)}
\prod_{\ell=2}^n(-\zeta_{1}^2-\ti\mu_\ell(\zeta_1)) 
\prod_{\substack{\ell=1\\ \ell\neq j}}^n(\ti\mu_j(\zeta_1)-
\ti\mu_\ell(\zeta_1))^{-1}, \lb{5.76} \\
& \hspace*{7.3cm}  j=2, \dots, n. \no
\end{align}
Since the right-hand sides in \eqref{5.75}, \eqref{5.76} are holomorphic
with respect to the $n$ variables $\zeta_1,\ti\mu_2,\dots,\ti\mu_n$ for
$\zeta_1$ near zero and $\ti\mu_j$ near $[E_{2j-1},E_{2j}]$,
$j=2,\dots,n$, equations \eqref{5.75} and \eqref{5.76} yield solutions
$x,\ti\mu_2,\dots,\ti\mu_n$ holomorphic with respect to $\zeta_1$ near
$\zeta_1=0$. In particular, since 
\begin{equation}
x'(\zeta_1)\underset{\zeta_1\to 0}{=}\zeta_1^2/C_1+\Oh(\zeta_1^3), \quad 
\mu_j'(\zeta_1)\underset{\zeta_1\to 0}{=}C_j\zeta_1^2+\Oh(\zeta_1^3), \;
j=2,\dots,n, \lb{5.77}
\end{equation}
for some constants $C_j\neq 0$, $j=1,\dots,n$, one obtains
\begin{align}
x(\zeta_1)&\underset{\zeta_1\to 0}{=}\xi_{k}+\zeta_1^3/(3C_1)
+\Oh(\zeta_1^4), \lb{5.78} \\
\ti\mu_j(\zeta_1)&\underset{\zeta_1\to 0}{=}\ti\mu_j(0)+C_j\zeta_1^3/3
+\Oh(\zeta_1^4)= \mu_j(\xi_{k})+C_j\zeta_1^3/3
+\Oh(\zeta_1^4). \lb{5.79}
\end{align}  
Thus, inverting $x(\zeta_1)$, one observes
\begin{align}
\zeta_1(x)&\underset{x\to\xi_{k}}{=}[3C_1(x-\xi_{k})]^{1/3} +
\Oh\big((x-\xi_{k})^{2/3}\big), \lb{5.80} \\
\mu_j(x)&\underset{x\to\xi_{k}}{=}\mu_j(\xi_{k})+C_1C_j(x-\xi_{k}) 
+ \Oh\big((x-\xi_{k})^{4/3}\big), \quad j=2,\dots,n, \lb{5.81} 
\end{align}
and hence,
\begin{align}
\mu_{1,x}(x)&=-2\zeta_{1}(x)\zeta_{1,x}(x) \no \\
& \hspace*{-2.5mm} \underset{x\to \xi_{k}}{=}
-(2/3)(3C_{1})^{2/3}(x-\xi_{k})^{-1/3}+
\oh\big((x-\xi_{k})^{-1/3}\big) \lb{5.82}
\end{align}
and  
\begin{align}
u_x(x)&=(1/2)\sum_{j=1}^n \mu_{j,x}(x) \no \\
& \hspace*{-2.5mm} \underset{x\to \xi_{k}}{=}
-(1/3)(3C_{1})^{2/3}(x-\xi_{k})^{-1/3}+
\oh\big((x-\xi_{k})^{-1/3}\big). \lb{5.83}
\end{align}
The singular behavior \eqref{5.82}, \eqref{5.83} repeats itself after each
revolution of $\mu_{1}$ around its circle and occurs whenever
$\mu_{1}$ passes again through $E_{1}=0$, giving rise to the 
exceptional set $\{\xi_k\}_{k\in\bbZ}$ in \eqref{5.64}. Hence, $\mu_j\in
C^1(\bbR)$, 
$j=2,\dots,n$, while $\mu_{1,x}=-\zeta_1(x)^2$ blows up whenever $x$
approaches an element of $\{\xi_k\}_{k\in\bbZ}$. The rest of the
discussion follows as in Theorem \ref{t5.8} and Corollary \ref{c5.9}. Since
$\mu_{1}(x_0)=E_{1}=0$ is not an admissible initial condition in
\eqref{5.63}, one point must be removed from the circle associated to
$\mu_{1}$ which topologically results in $\bbR$ instead of $S^1$ and
hence in the noncompact isospectral manifold $\bbT^{n-1}\times\bbR$.
\end{proof}

Thus, smooth algebro-geometric CH solutions require $E_0=0$ or
$E_{2n+1}=0$.

\vspace*{2mm}
{\bf Acknowledgments.}
We are indebted to Mark Alber, Roberto Camassa, Adrian Constantin, Yuri
Fedorov, and Igor Krichever for stimulating discussions. 


\end{document}